\documentclass[journal ]{new-aiaa}
\usepackage[utf8]{inputenc}
\usepackage{textcomp}

\usepackage{graphicx}
\usepackage{amsmath}
\usepackage[version=4]{mhchem}
\usepackage{siunitx}
\usepackage{longtable,tabularx}
\usepackage{float}
\usepackage{bm}
\usepackage{siunitx}
\usepackage{caption}
\usepackage{subcaption}
\usepackage[export]{adjustbox}
\usepackage{array}
\usepackage{tabularx}
    \newcolumntype{C}{>{\centering\arraybackslash}b{1.5cm}}
    \newcolumntype{D}{>{\centering\arraybackslash}b{1.55cm}}
\usepackage{booktabs}
\usepackage{multirow}
\usepackage[table,dvipsnames]{xcolor}% http://ctan.org/pkg/xcolor
\newcommand{\ubar}{\bm{\overline{u}}}
\newcommand{\red}{}
\newcommand{\blue}{}

\title{Simulations of Intracycle Angular Velocity Control\\ for a Cross-Flow Turbine}

\author{Mukul Dave\footnote{PhD student, Mechanical Engineering, AIAA Member.}}
\affil{University of Wisconsin-Madison, Madison, WI, 53715}

\author{Benjamin Strom\footnote{Co-Founder, XFlow Energy Company.}}
\affil{XFlow Energy Company, Seattle, WA, 98108}

\author{Abigale Snortland\footnote{PhD student, Mechanical Engineering.}}
\affil{University of Washington, Seattle, WA, 98195}

\author{Owen Williams\footnote{Research Assistant Professor, Aeronautics \& Astronautics, AIAA Senior Member.}}
\affil{University of Washington, Seattle, WA, 98195}

\author{Brian Polagye\footnote{Associate Professor, Mechanical Engineering.}}
\affil{University of Washington, Seattle, WA, 98195}

\author{Jennifer A. Franck\footnote{Assistant Professor, Engineering Physics, AIAA Senior Member.}}
\affil{University of Wisconsin-Madison, Madison, WI, 53715}

\begin{document}

\maketitle

\begin{abstract}
Straight-bladed cross-flow turbines are computationally explored for harvesting energy in wind and water currents. One challenge for cross-flow turbines is the transient occurrence of high apparent angles of attack on the blades that reduces efficiency due to flow separation. This paper explores kinematic manipulation of the apparent angle of attack through intracycle control of the angular velocity. Using an unsteady Reynolds-averaged Navier-Stokes (URANS) model at moderate Reynolds numbers, the kinematics and associated flow physics are explored for confined and unconfined configurations. The computations demonstrate an increase in turbine efficiency up to 54\%, very closely matching the benefits shown by previous intracycle control experiments. Simulations display the time-evolution of angle of attack and flow velocity relative to the blade, which are modified with sinusoidal angular velocity such that the peak torque generation aligns with the peak angular velocity. With optimal kinematics in a confined flow there is minimal flow separation during peak power generation, however there is a large trailing edge vortex (TEV) shed as the torque decreases. The unconfined configuration has more prominent flow separation and is more susceptible to Reynolds number, resulting in a 41\% increase in power generation under the same kinematic conditions as the confined flow.
\end{abstract}

\section*{Nomenclature}

%\noindent(Nomenclature entries should have the units identified)

{\renewcommand\arraystretch{1.0}
\noindent\begin{longtable*}{@{}l @{\quad=\quad} l@{}}
$A$ & projected area of the turbine normal to the flow  \\
$A_\omega$ & amplitude of sinsusoidal variation of $\omega$  \\
$c$ & chord length of the turbine blade  \\
$C_P$ & power coefficient or conversion efficiency of the turbine  \\
$\overline{C_P}$ & average power coefficient  \\
$C_Q$ & normalized value of torque or torque coefficient  \\
$k$ & turbulent kinetic energy  \\
$\overline{p}$ & mean component of pressure \\
$q$ & torque applied to the turbine by the fluid  \\ % changed tau to q as we use tau for the Reynolds stress tensor
$R$ & outermost radius of the turbine  \\
$Re$ & Reynolds number based on blade chord length and freestream velocity  \\
$\bm{U_\infty}$ & freestream velocity  \\
$\bm{U_n}$ & relative flow velocity (in the reference frame of the blade)  \\
${U_n^*}$ & non-dimensional magnitude of the relative flow velocity  \\
$\ubar$ & mean component of velocity vector\\
$\bm{V_\omega}$ & linear velocity of the blade  \\
$y^+$ & dimensionless wall distance  \\

$\alpha_n$ & apparent angle of attack of the flow (in the reference frame of the blade)  \\
$\alpha_p$ & preset pitch angle of the blade relative to the tangential direction of its motion  \\
$\Delta y$ & wall-normal distance \\
$\theta$ & azimuthal angular position of the blade  \\
$\lambda$ & tip speed ratio  \\
$\mathrm{\lambda_{ci}}$ & swirl strength  \\
$\nu$ & kinematic viscosity of the fluid  \\
$\rho$ & density of the fluid  \\
$\bm{\tau}$ & Reynolds stress tensor  \\
$\phi_\omega$ & phase shift between $\theta$ and sinusoidal variation of $\omega$  \\
$\Omega$ & vorticity \\
$\omega$ & angular velocity of the blade  \\
$\overline{\omega}$ & mean of sinsusoidal variation of $\omega$ 
\end{longtable*}}

\section{Introduction} \label{sec:intro}

Cross-flow or vertical-axis turbines convert energy in wind or flowing water to rotational mechanical energy via blades that rotate about an axis perpendicular to the flow. In traditional axial-flow turbines (AFTs) \cite{Kirke2011,AslamBhutta2012}, the rotational axis is perfectly aligned with the flow.  For cross-flow turbines, the axis of rotation is such that it does not require oncoming flow at a particular orientation. This alleviates the need for yaw control, which is critical for many AFT designs. Straight-bladed cross-flow turbines are also easy to manufacture due to less geometric complexity of the blade, and can potentially be quieter and safer than AFTs due to lower tip speeds. Furthermore, when oriented vertically, cross-flow turbines have the option to house the generator and mechanical components on the ground for harvesting wind energy, or at the water surface for hydrokinetic energy, decreasing maintenance costs. 

\begin{figure}
    \centering
    %\captionsetup[subfigure]{aboveskip=+12pt}
    \begin{subfigure}[b]{0.45\textwidth}
        \centering
        \includegraphics[width=\textwidth]{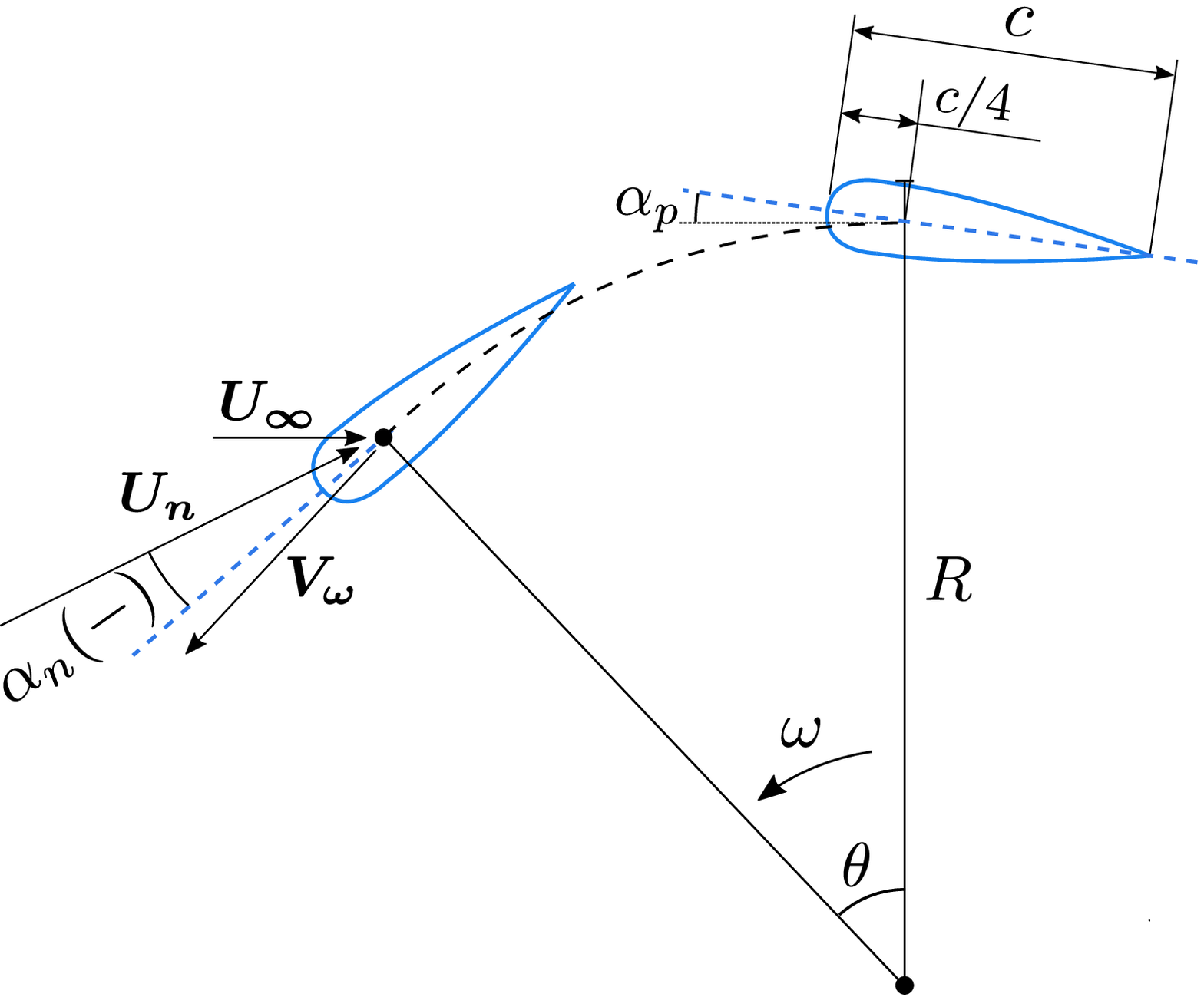}
        \caption{Schematic diagram and nomenclature of turbine blade kinematics.}
        \label{subfig:foil_kinematics_a}
    \end{subfigure}
    \hspace{1cm}
    %\captionsetup[subfigure]{aboveskip=0pt}
    \begin{subfigure}[b]{0.45\textwidth}
        \centering
        \includegraphics[width=\textwidth]{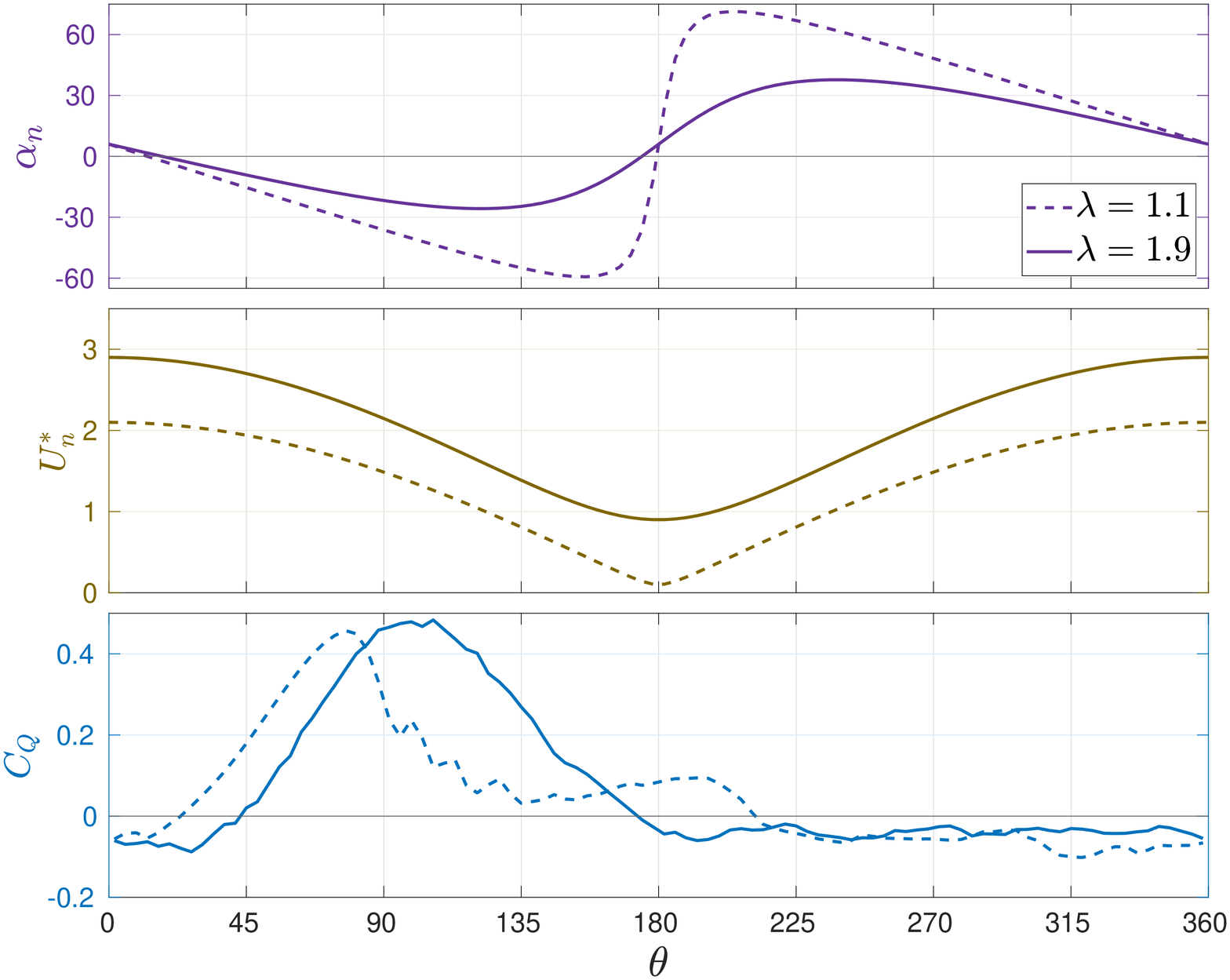}
        \caption{Variation of the apparent angle of attack, relative flow velocity and torque response for a single blade during rotation, $\alpha_p=6^\circ$ (URANS simulation for a two-bladed turbine in confined flow).}
        \label{subfig:foil_kinematics_b}
    \end{subfigure}
    \caption{}
    \label{fig:foil_kinematics}
\end{figure}

Fig. \ref{subfig:foil_kinematics_a} demonstrates the kinematics of a cross-flow turbine blade through a portion of its cycle. As the blade rotates in the freestream with an angular velocity, $\omega$, the instantaneous velocity of the blade is shown as $\bm{V_\omega}$. Lift is generated through the apparent angle of attack, $\alpha_n$, which is the angle the blade encounters with respect to the relative flow velocity, $\bm{U_n}$. The component of this lift force that is tangential to the blade motion drives the turbine.
Turbine performance is characterized as a function of tip speed ratio, defined as the ratio of turbine tip speed to the flow freestream velocity,

\begin{equation}
\label{equation:lambda}
\lambda(\theta) = \frac{\omega(\theta)R}{U_\infty} .
\end{equation}
\red{where $R$ is the distance from center of the turbine to the outermost edge of the blade as shown in Fig. \ref{subfig:foil_kinematics_a}.}
In this paper, $\theta=0^\circ$ represents the foil position when the velocity vector of the foil is directly opposing the freestream velocity vector.
Neglecting induced flow effects and the curvilinear nature of the relative flow, variation in $\alpha_n$ and $U_n^*$ throughout the rotation of a blade is given by

\begin{equation}
\label{equation:alpha_n}
    \alpha_n(\theta) = -\tan^{-1} \left( \frac{\sin \theta}{\lambda(\theta) + \cos \theta} \right) + \alpha_p(\theta)
\end{equation}

and 

\begin{equation}
\label{equation:u_nstar}
    U_n^* = \frac{|\bm{U_n}(\theta)|}{U_\infty} = \sqrt{\lambda(\theta)^2 + 2\lambda(\theta) \cos \theta + 1} .
\end{equation}
\red{where the sign of $\alpha_n$ is positive when the relative flow velocity is directed outwards from the axis of rotation and negative when it is directed inwards as in Fig. \ref{subfig:foil_kinematics_a}.} The torque applied to the turbine by the fluid and the resultant power generated are normalized respectively as the torque coefficient and power coefficient,

\begin{equation}
C_Q(\theta) = \frac{q(\theta)}{\frac{1}{2} \rho U_\infty^2 A R}\ 
\end{equation}

and

\begin{equation}
C_P(\theta) = \frac{q(\theta) \omega(\theta)}{\frac{1}{2} \rho U_\infty^3 A}
\end{equation}
where $A=2R$ for a two-dimensional model. The power coefficient, $C_P$, is the conversion efficiency of the turbine defined by the ratio of the power generated to the power available in the projected area.

Fig. \ref{subfig:foil_kinematics_b} shows the variation of $\alpha_n$ and $U_n^*$ during a rotation cycle for two tip speed ratios. Phase-averaged torque generation on a single blade, obtained through URANS simulations (one blade of a two-bladed turbine in confined flow), is also included as an example.
The apparent angle of attack undergoes very high variation during the cycle, reaching $-25^\circ$ to $+35^\circ$ for $\lambda=1.9$ and approximately +/- $60^\circ$ for $\lambda=1.1$, which is much higher than the critical angle of attack for flow separation on a stationary foil ($\approx 12^\circ$) \cite{Timmer2008}. \red{Although the computed angle of attack in Fig. \ref{subfig:foil_kinematics_b} does not account for the effect of induction, the high variation} creates flow transitions from attached flow to highly separated flow and vice versa, a phenomenon known as dynamic stall. 
Uncontrolled dynamic stall produces a drop in the lift-to-drag ratio, decreasing the torque generated. For the example turbine shown in Fig. \ref{subfig:foil_kinematics_b}, the sharp decrease in torque for $\lambda=1.1$ at $\theta = 80^\circ$ is coincident with the onset of dynamic stall.

Due to the challenges posed by dynamic stall in cross-flow turbines, it has been studied extensively through laboratory experiments in which the evolution of vortices on the blades has been visualized and analyzed. Generation of two pairs of stall vortices at the blade has been observed during rotation \cite{Fujisawa2001}, and the roll up and formation of a trailing edge vortex has also been analyzed in detail \cite{SimaoFerreira2009}. \red{Dunne \& McKeon \cite{Dunne2015} developed a lower-order model of dynamic stall by performing experiments on a sinusoidally pitching and surging foil, and extrapolated it to a cross-flow turbine to predict the onset and completion of flow separation. Although this captures the variation in relative velocity and angle of attack, the model does not take into account  flow curvature, induction, and the Coriolis force which are observed for the rotational motion of a turbine blade.}
Several computational fluid dynamics (CFD) studies have also been performed \cite{Buchner2015,SimaoFerreira2010,Wang2010,Tsai2016,Wang2016,Almohammadi2015,Lei2017,Hand2017,Lopez2016,Bachant2016,Ouro2018,Amet2009}. \red{The Coriolis effect has been shown to result in the turbine blade capturing a vortex pair at low tip speed ratios after the angle of attack starts decreasing \cite{Tsai2016}. This negatively affects torque generation and cannot be reproduced by a non-rotational pitching-surging blade with an equivalent angle of attack and relative velocity variation.} Wang et al. observed how vortex dynamics plays a role in reduction of power conversion at both low and high tip speed ratios \cite{Wang2016}. The effect of blade shape on the dynamic stall process, simulated with a prescribed pitching motion, has shown that a cambered blade (NACA4412) can perform better than a symmetric blade (NACA0012) due to higher lift-to-drag ratio and a slight delay in dynamic stall \cite{Ouro2018}.

In attempts to mitigate dynamic stall, flow control techniques may be used to subdue flow separation on the turbine blades. Passive control solutions such as serrations on the leading edge \cite{Wang2017}, or a cavity or dimple on the suction surface \cite{Sobhani2017} have been performed, showing moderate improvement by suppressing or delaying flow separation.
Active control such as oscillating flaps \cite{Xiao2013}, synthetic jets \cite{Yen2013,Velasco2017}, plasma actuators \cite{Greenblatt2012}, and boundary layer suction \cite{Morgulis2016} have also been investigated, often yielding more enhancement in power than passive control, but these mechanisms are costly to integrate and maintain.

Dynamic stall can also temporarily generate high lift due to the leading edge vortex (LEV) formed when the boundary layer separates from the foil at moderate $\alpha_n$ \cite{Franck2017, Ribeiro2020, Keisar2020}. However this benefit is temporary, and there is a loss in lift as the vortex is shed. Thus, careful control of the $\alpha_n$ profile is an approach to control the LEV and other unsteady flow mechanisms associated with the dynamic stall process. 

Equation \ref{equation:alpha_n} demonstrates how variation of the angle of attack during circular rotation can be controlled through control of either $\alpha_p(\theta)$ or $\lambda(\theta)$.
Intracycle variation of the blade pitch angle, $\alpha_p$, has been previously explored through design and theoretical performance analyses \cite{Schonborn2007,Paraschivoiu2009,Lazauskas1992}, experiments \cite{Kirke2011a,Elkhoury2015} and simulations
\cite{RemiGosselin2013,Abdalrahman2017,Paillard2015,Elkhoury2015}. Most of these investigations indicate a maximum enhancement in turbine power of at least 25\%, especially for low tip speed ratio values.
However, both active and passive pitch control require additional complex mechanisms to be integrated in the design, which drives up turbine cost.

This paper investigates an alternative method proposed by Strom et al. \cite{Strom2017} who implemented intracycle variation of the angular velocity, $\omega$ (and hence, $\lambda$) to change the angle of attack and relative flow velocity profiles in Eqns. \ref{equation:alpha_n} and \ref{equation:u_nstar}. Strom et al. performed flume experiments at a Reynolds number of $Re = {cU_\infty}/{\nu} = 3.1\times10^4$ with an optimization process to determine the mean, amplitude, and phase-shift of the sinusoidal variation of $\omega$. They concluded that controlling $\alpha_n$ delays the drop in lift-to-drag ratio while the high torque generation phase coincides with high $\omega$, resulting in up to 53\% increase in average power generation.
\red{In actual deployment, the use of this method mandates a real-time control of the intracycle variation of $\omega$ based on the changes in oncoming flow speed and direction. However, the advantage of not requiring a yaw mechanism is maintained.}

Strom et al. further hypothesized that the controlled $\alpha_n$ and $U_n^*$ cycles created a sustained leading edge vortex (LEV) that enhanced lift generation on the foil, however this has not yet been confirmed through physical experiments. Experimentally, time-resolved flow visualization is challenging, and although more time-consuming, computational analysis can provide insight into the unsteady fluid mechanics, particularly close to the blade surface. 

Thus, the goal of this paper is a computational investigation into the physical mechanisms and vortex dynamics that are affected by altering the intracycle angular velocity. This is performed via unsteady Reynolds-averaged Navier-Stokes (URANS) simulations of straight-bladed cross-flow turbines under similar blockage and Reynolds number conditions as Strom et al. Additionally, simulations are also performed for an unconfined flow configuration with minimal blockage \red{to investigate the effects of confinement, which will vary with deployment setting (e.g., unconfined for a wind deployment).}
The computational methods are explained in the next section. A mesh sensitivity analysis and comparison with experimental results for validation is included in section \ref{sec:mesh}. In section \ref{sec:results}, performance results for cross-flow turbines with a sinusoidal variation of $\omega$ are presented. Experimental findings are compared with the computations and the vortex dynamics are thoroughly explored, including visualization of dynamic stall and vortex-structure interactions.

\section{Computational Methods} \label{sec:methods}

The computations utilize an incompressible URANS model whose governing equations are  

\begin{equation}
    {\nabla} \cdot \ubar = 0
    \label{eqn:rans_cont}
\end{equation}

\begin{equation}
    \frac{\partial \ubar}{\partial t} + (\ubar \cdot {\nabla}) \ubar = 
        - {\nabla} \overline{p} + \nu {\nabla}^2 \ubar - {\nabla} \cdot \bm{\tau}.
    \label{eqn:rans_mom}
\end{equation}

These equations are solved using a second-order accurate finite volume, pressure-implicit split-operator (PISO) method \citep{issa1986solution} implemented in {\it OpenFOAM} \citep{weller1998tensorial}. The Reynolds stress tensor, $\bm{\tau}$, in Eqn. \ref{eqn:rans_mom} is modeled with the {\it k-$\omega$} SST equations \citep{menter1994two}.
The {\it k-$\omega$} SST model has been used extensively in URANS simulations of cross-flow turbines \cite{Balduzzi2016,Paillard2015,Hand2017,RemiGosselin2013,Wong2018,Rogowski2018,Castelli2010} and is found to be a desirable closure model due to its documented ability of handling separated flows \cite{bardina1997turbulence}.
However, as with all models, it has been shown to have limitations, including lower fidelity solutions at high angles of attack \cite{Wang2010,Li2013} and in transitional Reynolds numbers \cite{Bianchini2017,Almohammadi2015}.

\begin{figure}[h]
    \centering
    \includegraphics[width=0.64\textwidth]{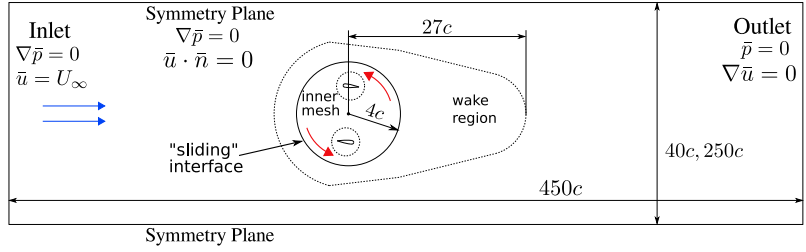}
    \caption{Schematic representation of the computational domain. \red{$c$ is the chord length of the foil, as shown in Fig. \ref{subfig:foil_kinematics_a}, and $c/R=0.47$.}}
    \label{fig:schematic}
\end{figure}

The simulation domain contains an inner mesh of radius $4c$ with a sliding interface between this rotating inner region and the intermediate wake-region mesh that remains stationary (Fig. \ref{fig:schematic}).
The wake region extends $27c$ downstream of the turbine center, while the outermost region of the domain extends $250c$ downstream and $200c$ upstream of the turbine center. Two different blockage ratios (ratio of projected frontal area of the turbine to total cross-section area of the domain) were investigated. For a domain width of $40c$, the resulting blockage is 10.6\% which closely mimics the setup for experiments compared with the simulations \cite{Snortland2019,Strom2017}. 
This configuration is referred to as the \textit{confined} configuration. 
\red{However, the flow is confined only across the turbine width for the two-dimensional computations, whereas it is confined along both turbine width and height in experiments, and hence the comparison of blockage ratios is approximate.}
Due to the relatively high blockage ratio of the confined configuration, and the resulting increase in power generation \cite{Bianchini2017,Lopez2016,Kinsey2017}, the performance of intracycle variation of $\omega$ was also tested with an \textit{unconfined} configuration that has a width of $250c$ corresponding to a blockage ratio of 1.7\%. A constant inlet flow velocity is prescribed on the left boundary while a constant pressure value is prescribed at the outlet boundary. A symmetry boundary condition is prescribed at the top and bottom that enforces zero normal flow velocity through the boundaries. A zero-gradient boundary condition is applied to the turbulent kinetic energy at the foil.
A wall function is utilized to calculate the specific dissipation for the first layer of mesh cells at the foil, that implements a blending function of its viscous sub-layer and log-law region variations based on distance from the wall \cite{Menter2001}.

For conditions where the flow separates from the blade, such as for the unconfined configuration, the results are particularly sensitive to the prescribed value of the far-field turbulent kinetic energy, $k$, which is used as a tuning parameter when benchmarking against experimental results.
There is no data readily available for unconfined configurations of this particular turbine. Barnsley and Wellicome's blockage correction \cite{barnsley1990final} has been found to yield good predictions of power generation in unconfined flow from experimental results for confined flow \cite{Ross2020}.
This correction is applied to the experimental data, and prescribing a far-field value of $k=1\times 10^{-5}$ yields 
a power coefficient within 6.6\% of the predicted blockage-corrected value for $\lambda=1.9$.

The current computational investigation is designed to mimic the experimental setup with the exception that it has no support structures nor center drive shaft. It is comprised of two NACA0018 foils with $c/R=0.47$, $\alpha_p = 6^\circ$ (leading edge angled outward), and $Re = {cU_\infty}/{\nu} = 4.5\times10^4$.

For intracycle variation of angular velocity, $\omega$ is made to vary sinusoidally with azimuthal position, given by

\begin{equation}
\omega(\theta) = \overline{\omega} + A_\omega sin(2\theta + \phi_\omega) .
\label{eqn:omega_intra}
\end{equation}
The frequency of $\omega$ variation is twice the rotation rate for a two-blade turbine.
The performance results for a cross-flow turbine rotating at constant $\omega$ with ${\lambda}=1.9$ are used as the baseline for comparison as the maximum power extraction occurs in this range of tip speed ratio for the confined flow configuration \cite{Snortland2019}. Hence for the purpose of this investigation, $\overline{\omega}$ is fixed at a value corresponding to ${\lambda}=1.9$. The amplitude is set as $A_\omega=0.63\ \overline{\omega}$, close to the optimal value arrived at by Strom et al. \cite{Strom2017}. 
For the computations, a manual optimization process leads to a phase shift of $\phi_\omega= \SI{3.0}{\radian}$ for maximum enhancement in power generation. Although the mean and amplitude of sinusoidal variation were optimized for a confined flow by Strom et al., the unconfined simulations are performed with the same parameters. 
In practice, power needs to be applied externally to accelerate and decelerate the rotor in order to generate a sinusoidal variation of $\omega$. This input power is not incorporated in the power coefficient here as it integrates to zero over each phase period \cite{Strom2017,Polagye2019}.

\section{Mesh Sensitivity and Validation with Experiments} \label{sec:mesh}

\begin{figure}[h]
    \centering
    \includegraphics[width=0.9\textwidth]{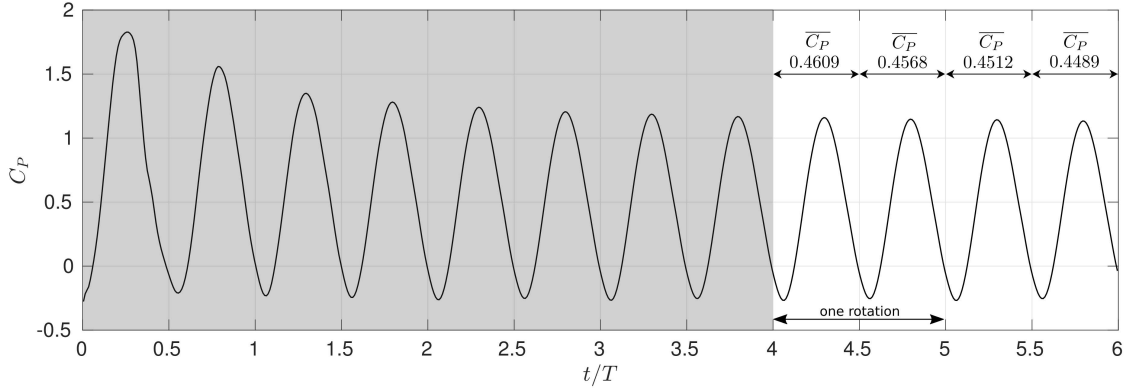}
    \caption{Evolution of power coefficient over six revolutions and twelve phase periods for $\lambda=1.9,\ Re = 4.5\times10^4$. The grey background indicates the developing flow state, and the white background indicates developed flow or steady state over which the power coefficient is averaged.}
    \label{fig:cp_steady}
\end{figure}

Simulations with a constant $\omega$ corresponding to $\lambda=1.9$ are performed for the mesh-sensitivity analysis. Fig. \ref{subfig:foil_kinematics_b} shows the variation of $\alpha_n$ and $U_n^*$ during a rotation cycle. Before rotation, a fully-developed state is achieved by running the simulation with stationary blades. The rotating simulation is then run for six rotor revolutions. Figure \ref{fig:cp_steady} shows the variation of power coefficient for the confined flow within these six complete revolutions. The force and moment values are phase-averaged over the last four half-revolutions to get the mean values once a fully developed state is reached (one phase period equals a half-revolution due to the symmetry of a two-blade turbine).

\begin{table}[!h]
\begin{center}
\caption{Details of different mesh configurations demonstrating mesh sensitivity for the URANS simulations. The highlighted meshes are selected for the simulations. $(\lambda=1.9,\ Re = 4.5\times10^4)$}
\begin{tabularx}{0.95\textwidth}{C C C D D D C C} 
\toprule
 &  & $\Delta y/c$ for first mesh layer & No. of body-fitted layers & No. of cells in the inner mesh & Total no. of cells & Reynolds number  & $\overline{C_P}$ \\ \midrule
 
 & {\color[HTML]{036400} mesh A} & {\color[HTML]{036400} 0.001}  & {\color[HTML]{036400} 60} & {\color[HTML]{036400} 178,077} & {\color[HTML]{036400} 272,892} & {\color[HTML]{036400} $4.5 \times 10^4$} & {\color[HTML]{036400} 0.454} \\

 & mesh B  & 0.001 & 60 & 318,837 & 400,646 & $4.5 \times 10^4$ & 0.437 \\

 & mesh C & 0.0005 & 60 & 316,017 & 420,352 & $4.5 \times 10^4$ & 0.441 \\

\multirow{-4}{*}{confined} & \multicolumn{5}{l}{\ \ Experimental data} & $4.5 \times 10^4$  & 0.377 \\ \midrule

 & mesh D & 0.001 & 60 & 223,357 & 307,300 & $4.5 \times 10^4$ & 0.228 \\

 & {\color[HTML]{036400} mesh E} & {\color[HTML]{036400} 0.001} & {\color[HTML]{036400} 60} & {\color[HTML]{036400} 318,837} & {\color[HTML]{036400} 427,870} & {\color[HTML]{036400} $4.5 \times 10^4$} & {\color[HTML]{036400} 0.324} \\

 & mesh E &  &  &  &  & $1 \times 10^5$  & 0.430 \\
 
 & mesh F & 0.001 & 60 & 370,409 & 535,720 & $4.5 \times 10^4$ & 0.346 \\

\multirow{-5}{*}{unconfined} & \multicolumn{5}{l}{\ \ Experimental data with blockage correction} & $4.5 \times 10^4$ & 0.305 \\ \bottomrule

\label{tab:meshes}
\end{tabularx}
\end{center}
\end{table}

\begin{figure}[!ht]
    \centering
    \includegraphics[width=0.5\textwidth]{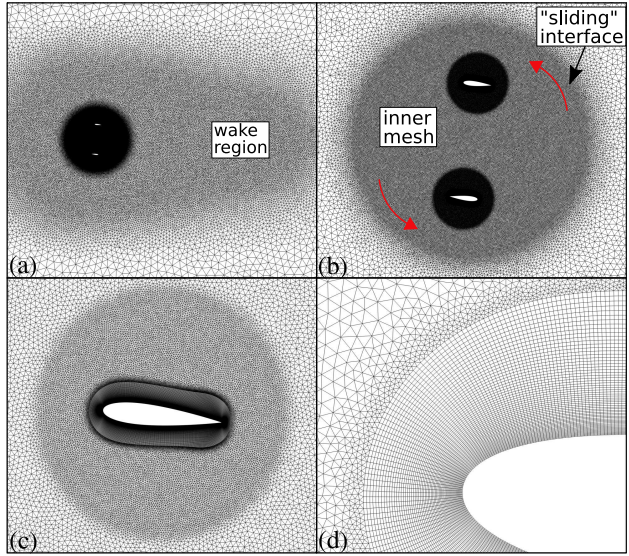}
    \caption{Layers of computational mesh with increasing level of resolution: (a) Immediate surrounding of the inner mesh (wake region). (b) Rotating inner mesh. (c) Mesh around the foil. (d) Body-fitted structured mesh to resolve the boundary layer.}
    \label{fig:mesh}
\end{figure}

Simulations are performed with various mesh configurations and the results are directly compared with experimental data to determine an appropriate computational setup. Results for six mesh configurations are presented in Table \ref{tab:meshes}. %The final mesh for the confined flow,
Mesh A is shown at different levels of resolution in Fig. \ref{fig:mesh}. The boundary layer region around the foil is resolved with layers of body-fitted structured mesh elements (Fig. \ref{fig:mesh}d) which transition to unstructured elements with gradually decreasing levels of refinement. For all mesh configurations, the number of mesh points along the foil boundary is 523, the wall-normal distance of the first mesh layer, $\Delta y / c$, is given in Table \ref{tab:meshes}, \red{and the mesh points within the boundary layer increase by a growth factor $\leq 1.1$.}

Table \ref{tab:meshes} tabulates the mean power coefficient for each of these meshes over the last four phase periods. The power coefficient and normalized forces vary minimally between the meshes A, B, and C, hence mesh A is chosen for performing the confined flow simulations. Mesh E is chosen for the unconfined simulations as the power and forces computed don't vary significantly between mesh E and mesh F.
For mesh A and mesh E, the maximum value of $y^+$ for the first mesh layer at the foil, for a stationary foil with $6^\circ$ angle of attack is 1.76 and 1.61 respectively.

\begin{figure}[t]
    \centering
    \includegraphics[height=1.48in]{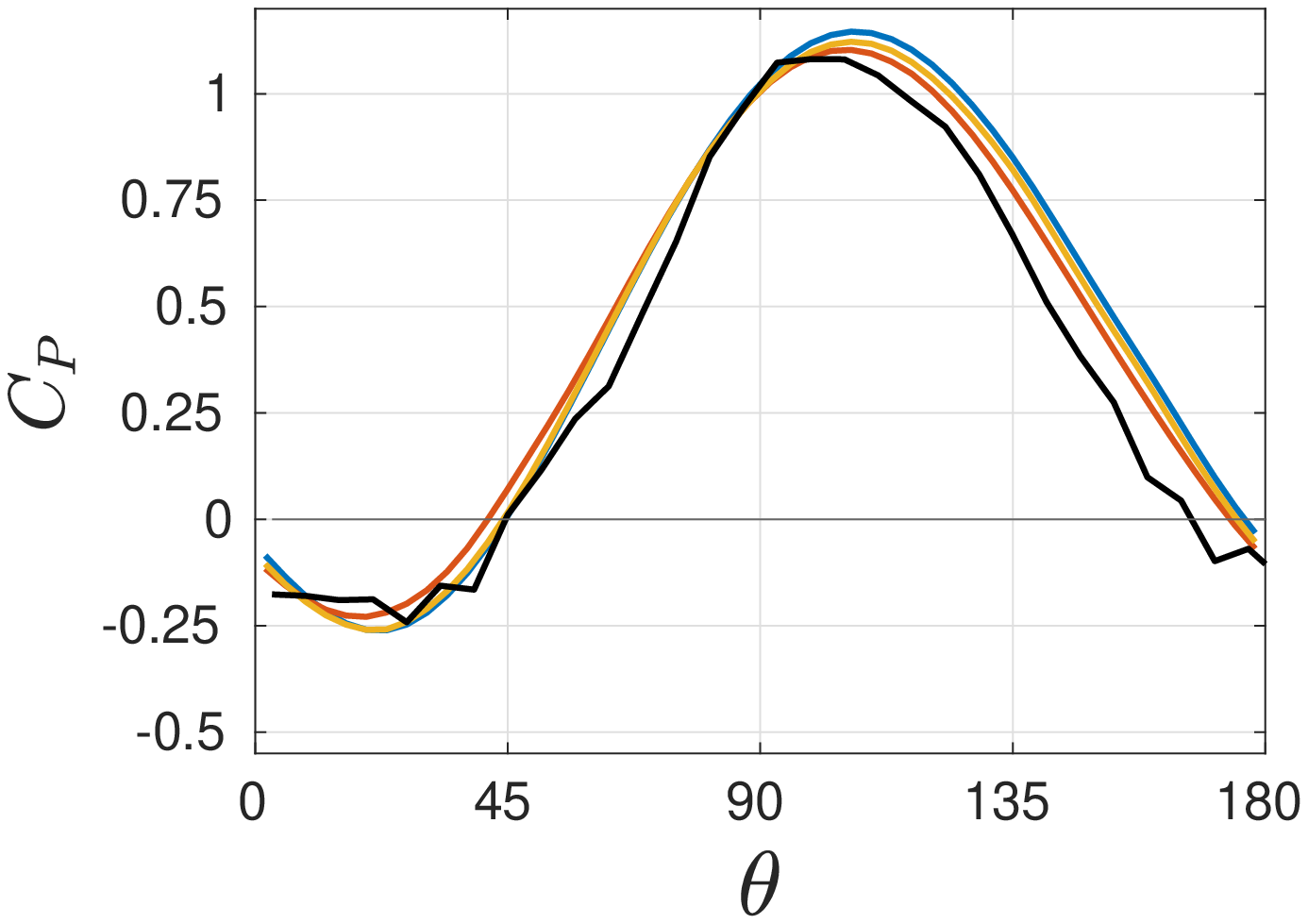} \hfill
    \includegraphics[height=1.48in]{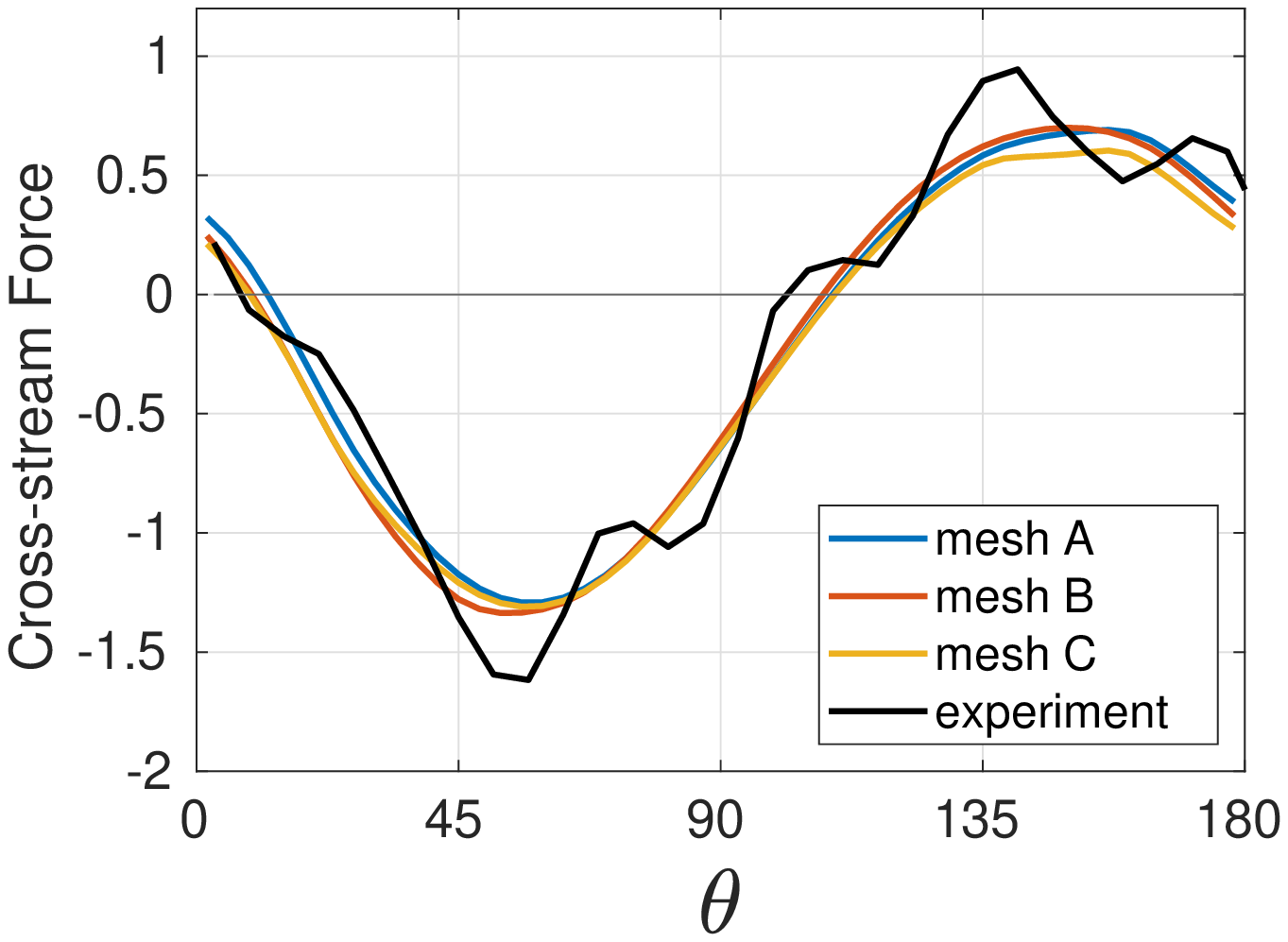} \hfill
    \includegraphics[height=1.52in]{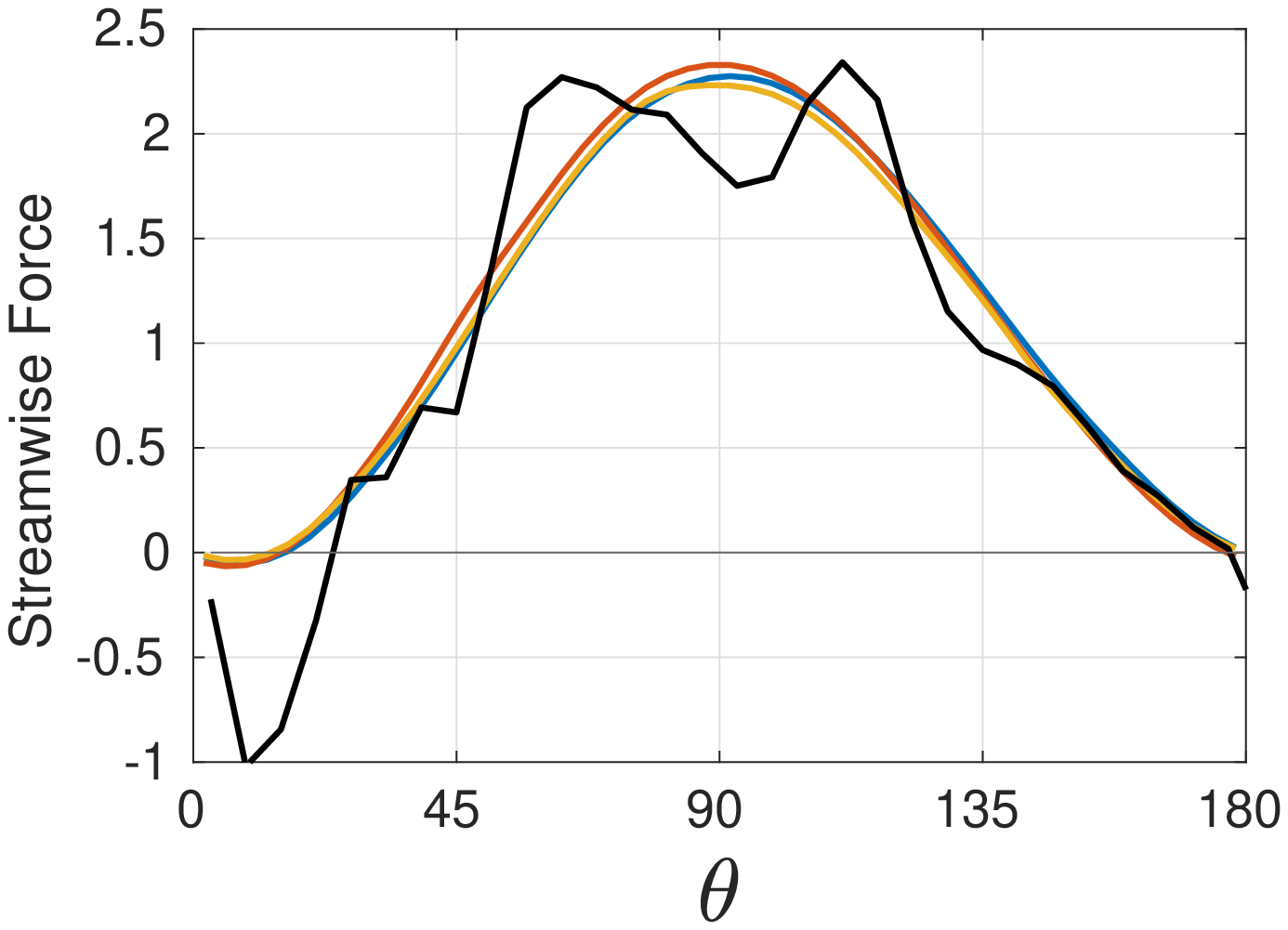}
    \caption{Power coefficient and force coefficients from URANS simulations for three different mesh configurations for confined flow (blockage ratio = 10.6\%), and from experimental data (blockage ratio = 11.6\%) for $\lambda=1.9,\ Re = 4.5\times10^4$. Mesh A is selected for the final analysis with confined flow.}
    \label{fig:mesh_validation}
\end{figure}

The computed power coefficient and force coefficients for the confined flow are shown in Fig. \ref{fig:mesh_validation} and directly compared with experiments by Snortland et al. \cite{Snortland2019},
\red{which are conducted at similar blockage ratio and Reynolds number to the computations. Streamwise and cross-stream forces are not reported in \cite{Snortland2019}, but acquired simultaneously by the 6-axis load cells mounted at either end of the turbine rotor and included here to augment validation.}
\red{As the simulations do not include support structures, the torques and forces on the blades were estimated from the experimental data. Specifically, the torques and forces arising from rotating only the center shaft and struts at the same inflow velocity were subtracted from those for the full turbine. This method has been demonstrated to accurately recover blade torque \cite{Strom2018,Bachant2016}. However, there is some uncertainty in the streamwise force on the support structure owing to the lack of induction and bypass flow in experiments when the blades are not present.}
The confined flow simulations show good agreement with experiments while over-predicting the peak power coefficient. 
There are fluctuations present in the experimentally measured forces which are likely associated with mechanical and hydrodynamic resonance of the experimental setup, however the values about which they fluctuate match with the computations.

\begin{figure}[!ht]
    \centering
    
    \begin{subfigure}{\textwidth}
        \centering
        \includegraphics[height=1.55in]{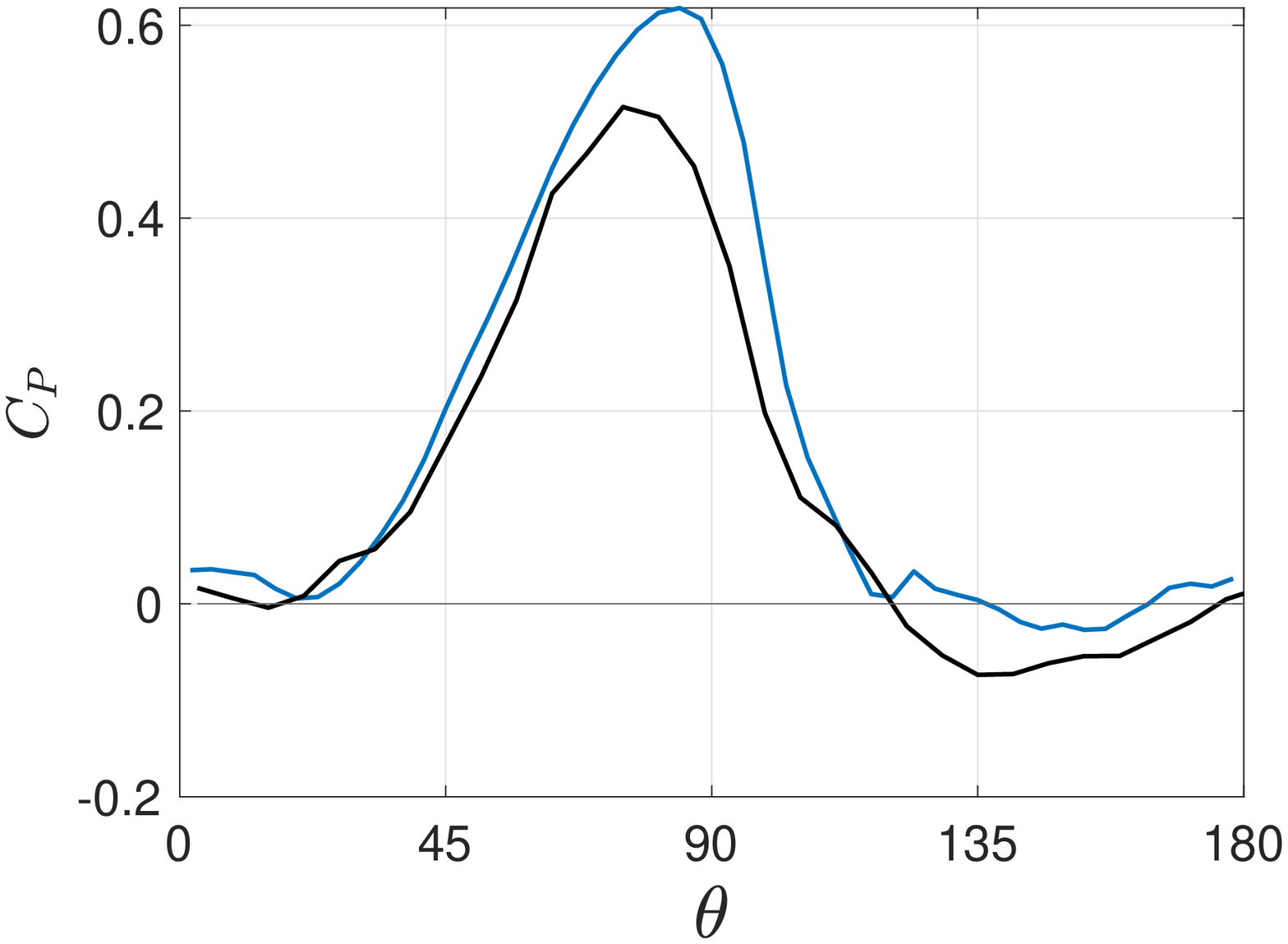} \hfill
        \includegraphics[height=1.55in]{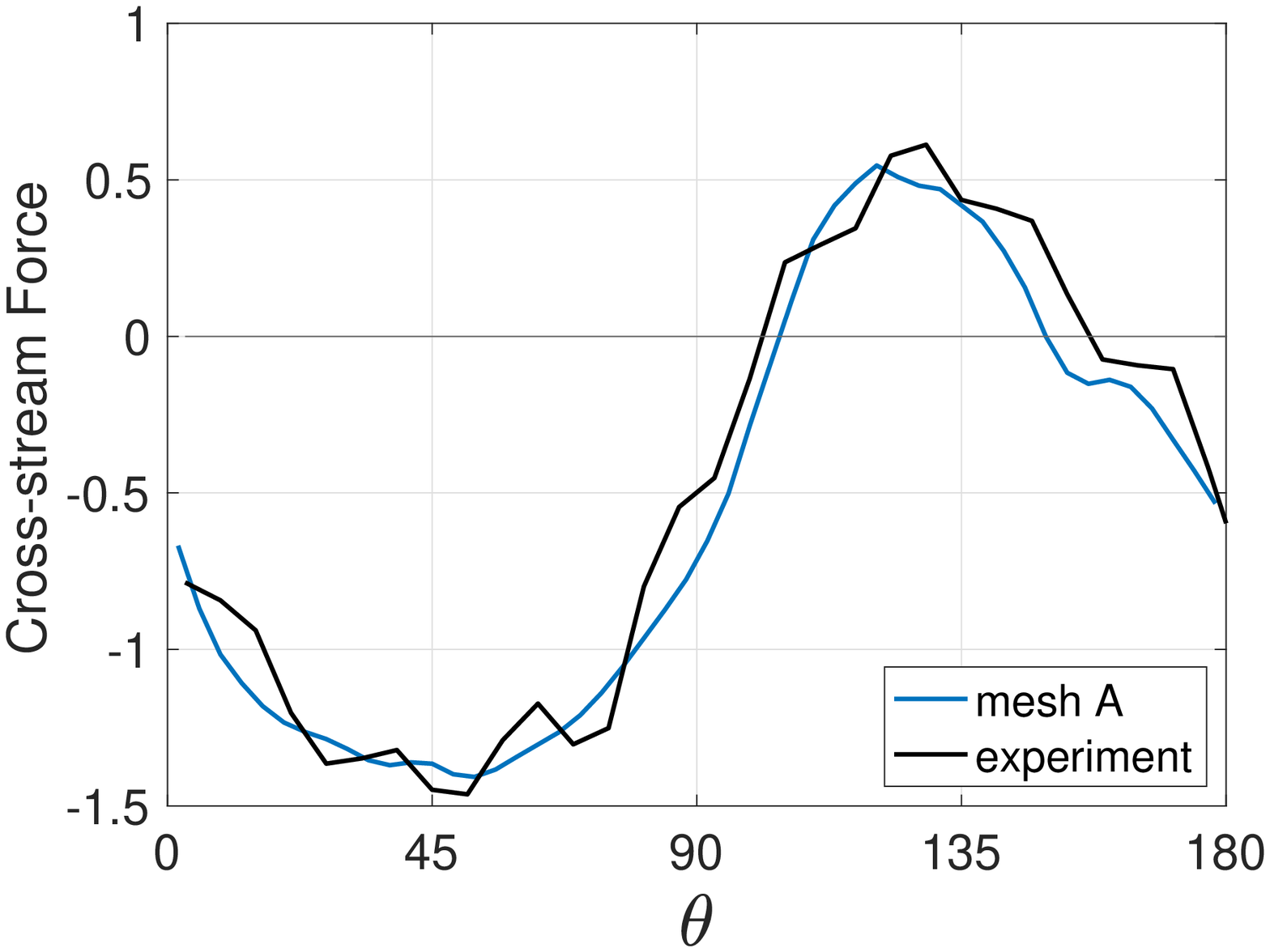} \hfill
         \includegraphics[height=1.55in]{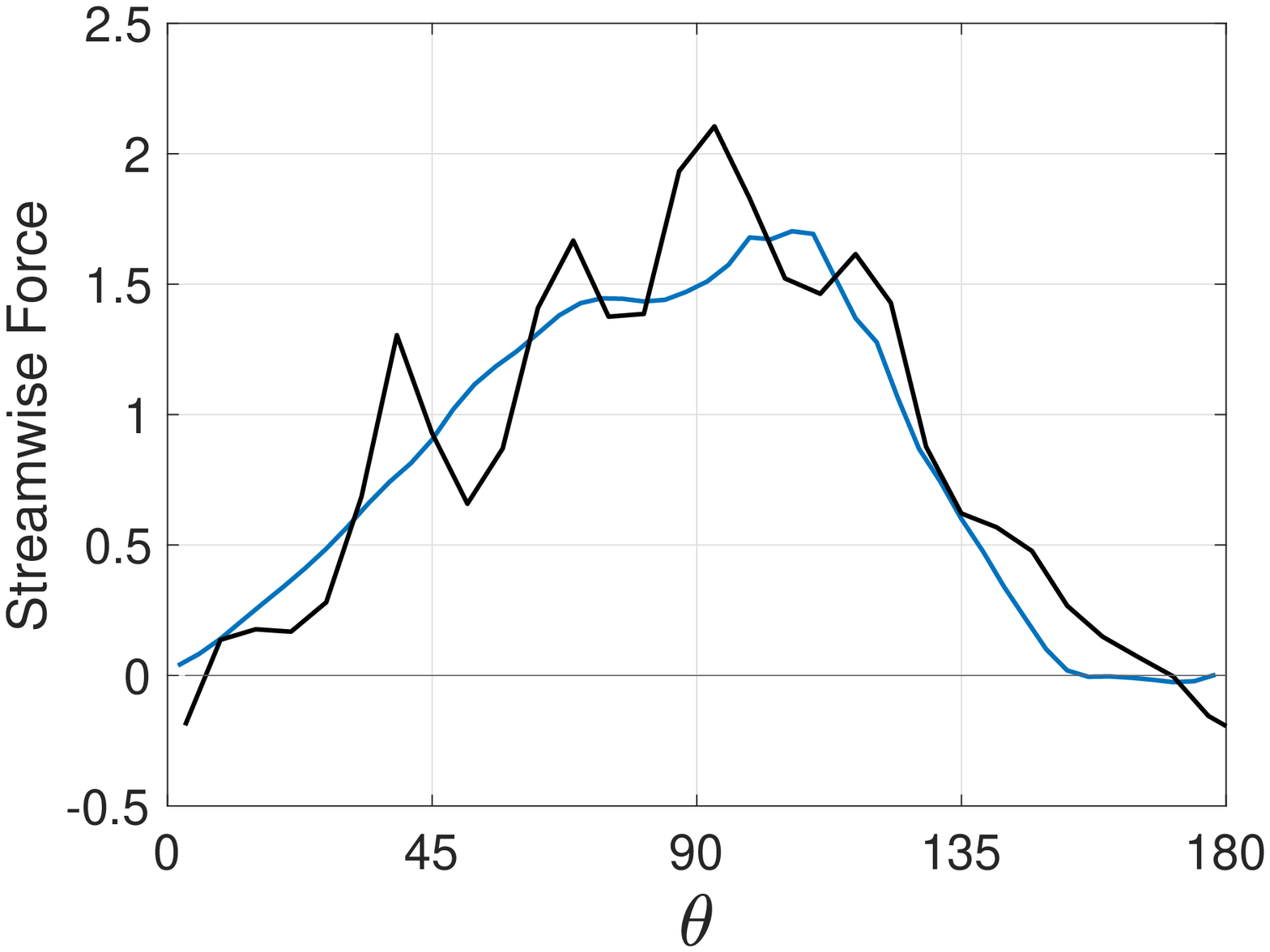}
        \caption{Power coefficient and force coefficients from URANS simulations (blockage ratio = 10.6\%), and from experimental data (blockage ratio = 11.6\%) for $\lambda=1.1,\ Re = 4.5\times10^4$.}
        \label{subfig:1_1_val}
    \end{subfigure}
    \vspace{0.4cm}
    
    \begin{subfigure}{\textwidth}
        \centering
        \setlength{\fboxsep}{0pt}%
        \setlength{\fboxrule}{.1pt}%
        \includegraphics[width=0.3\textwidth]{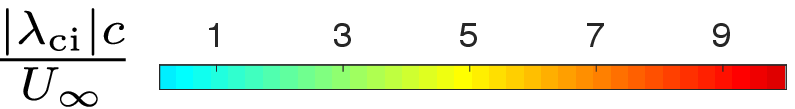} \hspace{0.1cm}
        \vspace{0.2cm}
        
        \fbox{\includegraphics[height=1.25in]{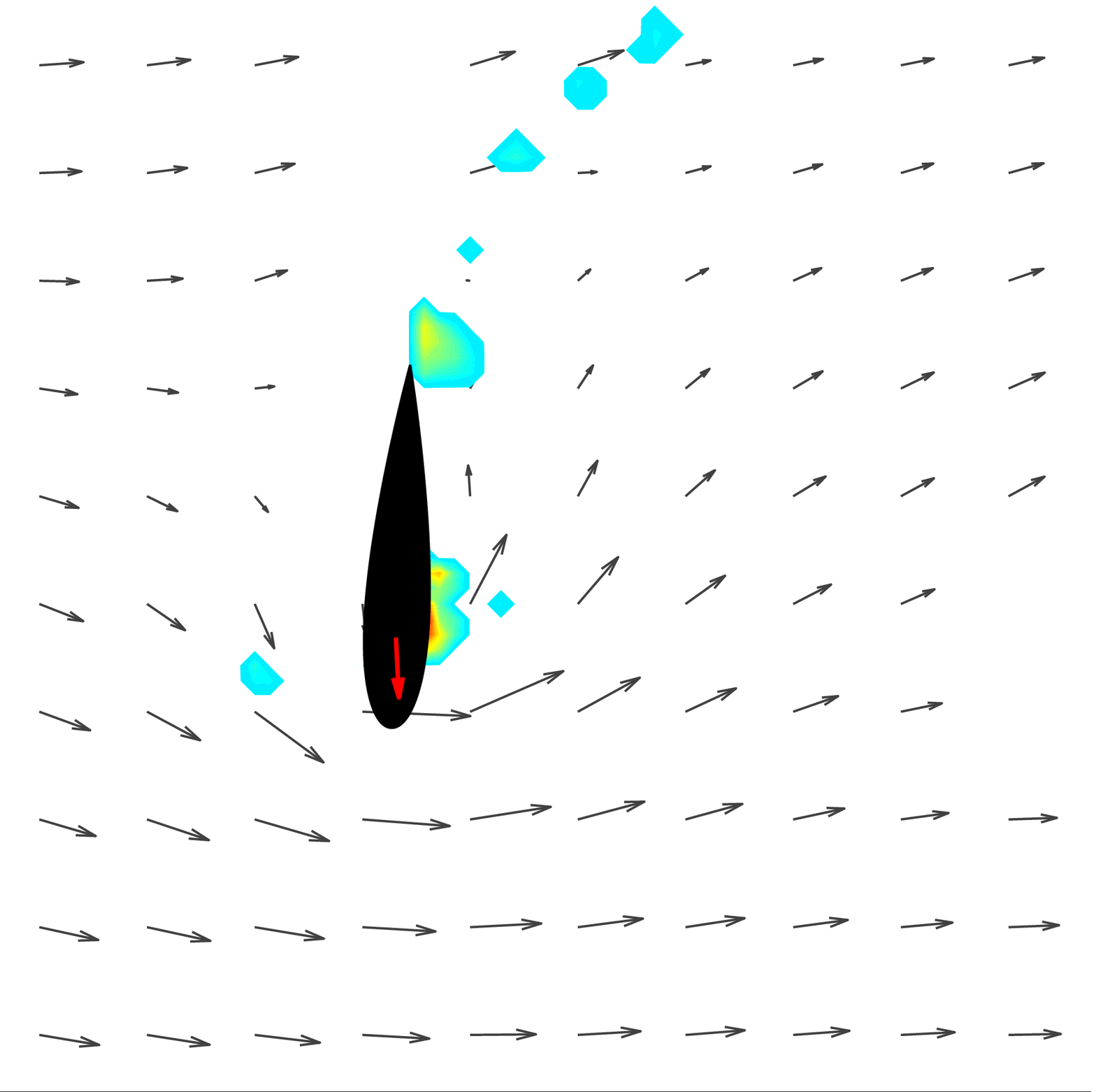}} \hfill
        \fbox{\includegraphics[height=1.25in]{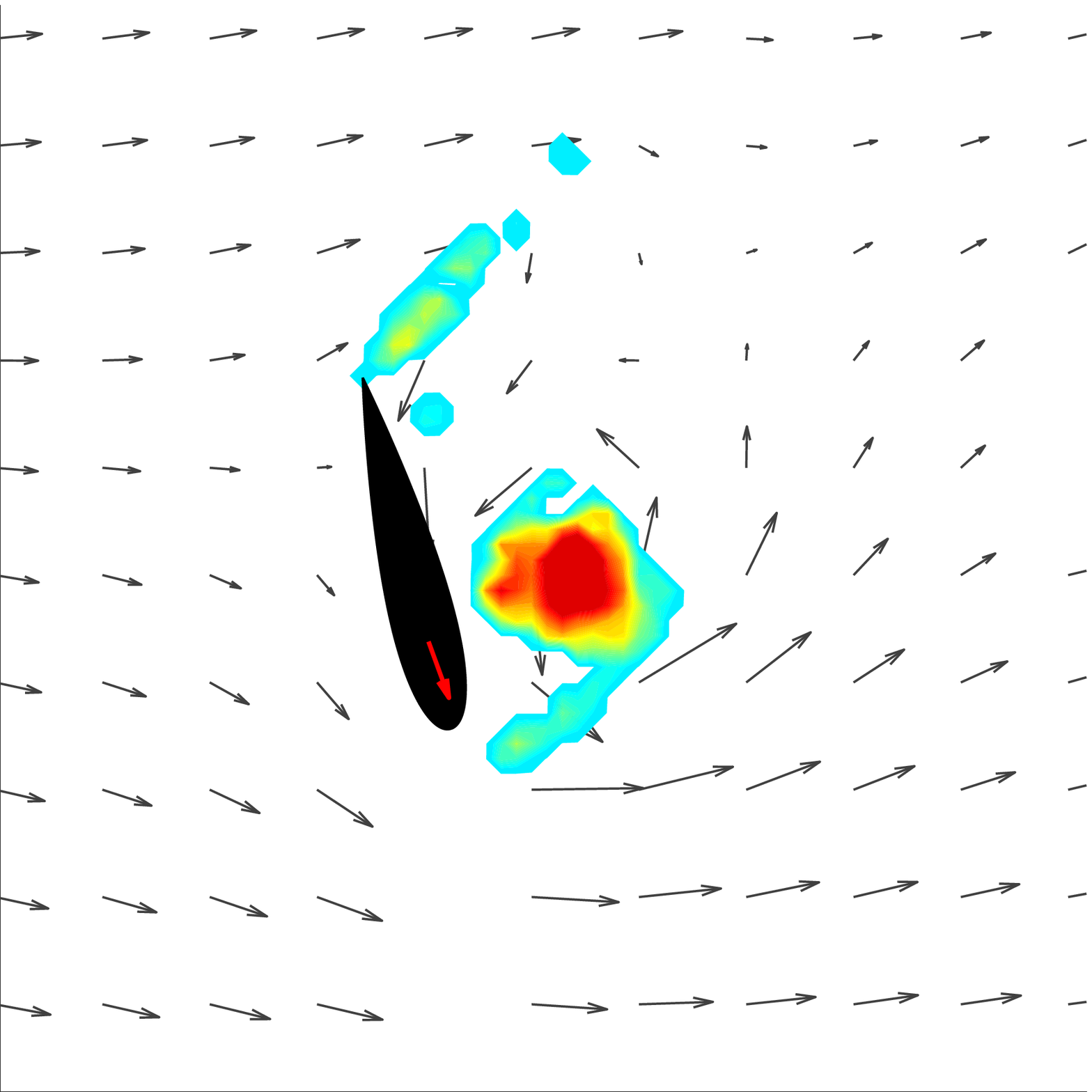}} \hfill
        \fbox{\includegraphics[height=1.25in]{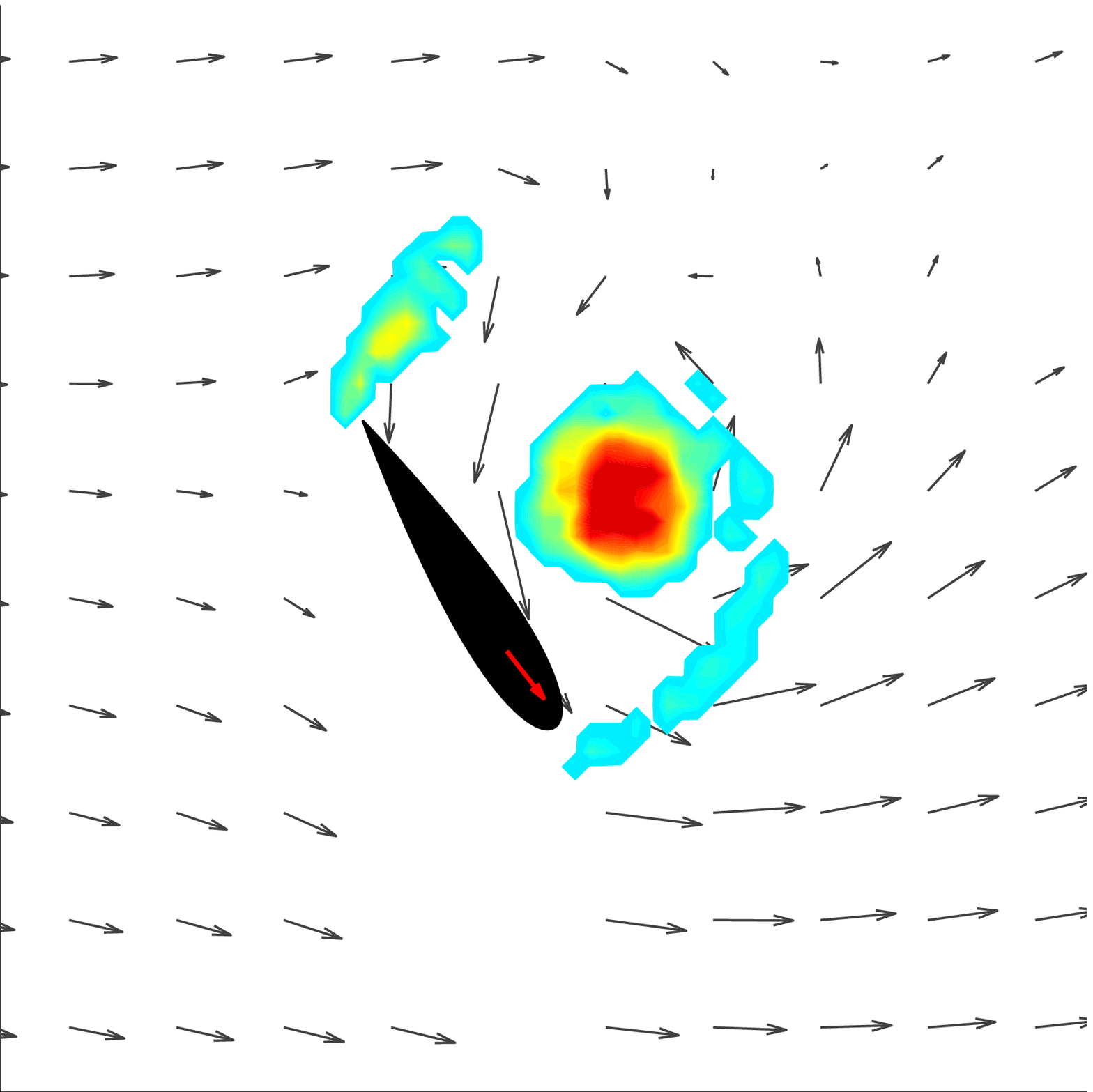}} \hfill
        \fbox{\includegraphics[height=1.25in]{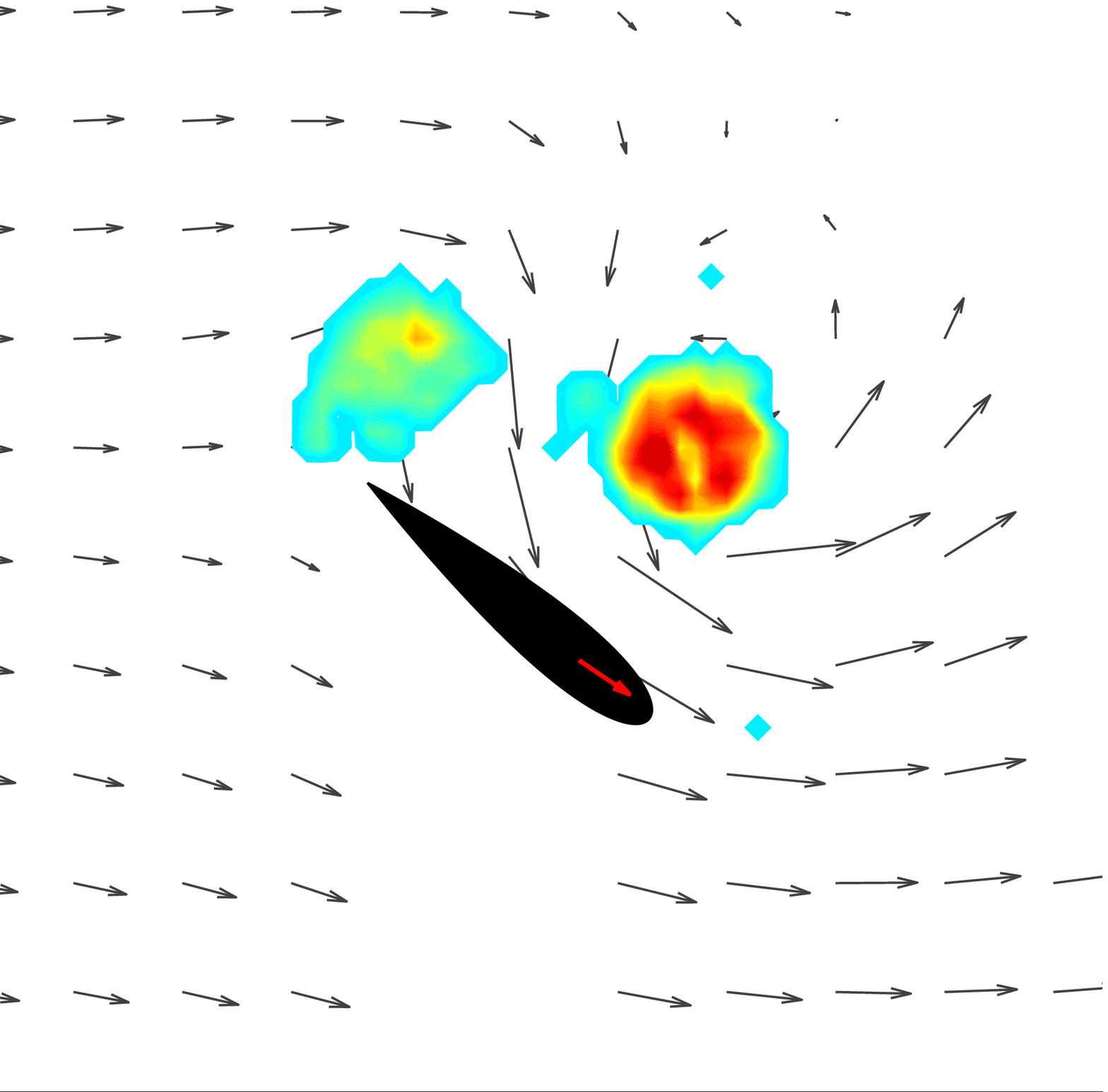}} \hfill
        \fbox{\includegraphics[height=1.25in]{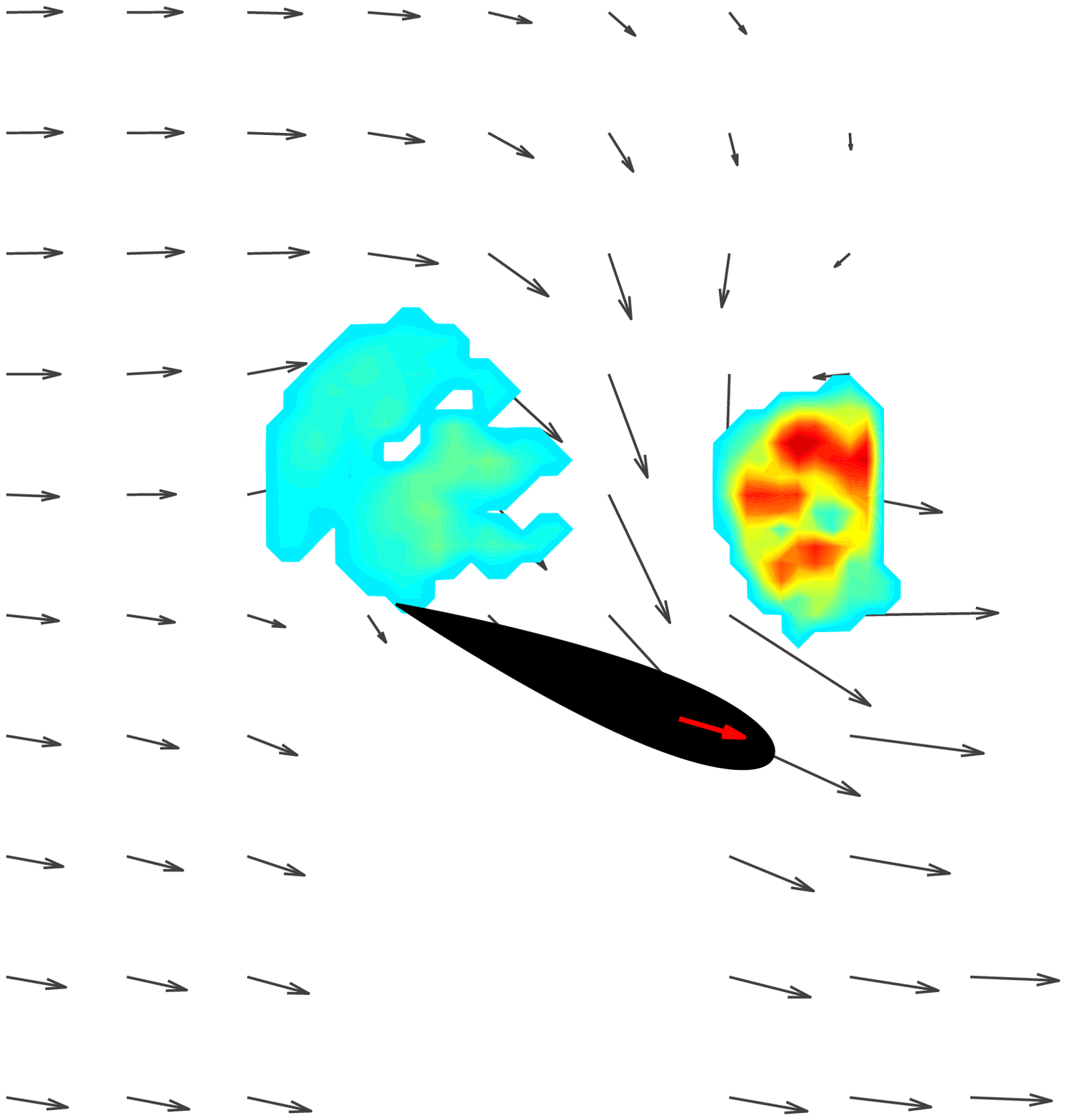}}
        \caption{Experimental data - PIV, blockage ratio = 11.6\%}
        \label{subfig:exp_ff}
    \end{subfigure} \vspace{0.1in}
    
    \begin{subfigure}{\textwidth}
        \centering
        \setlength{\fboxsep}{0pt}%
        \setlength{\fboxrule}{.1pt}%
        \fbox{\includegraphics[height=1.25in]{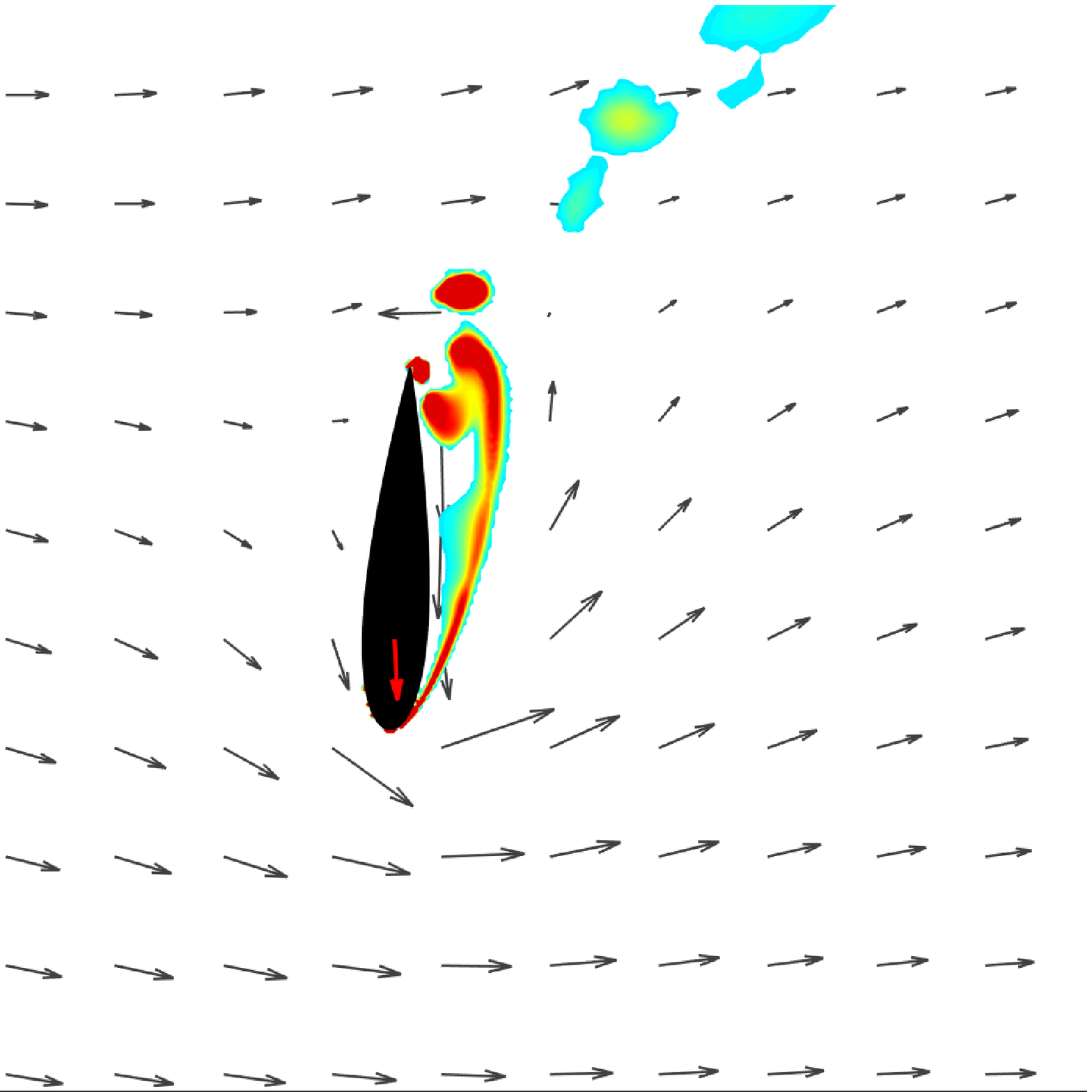}} \hfill
        \fbox{\includegraphics[height=1.25in]{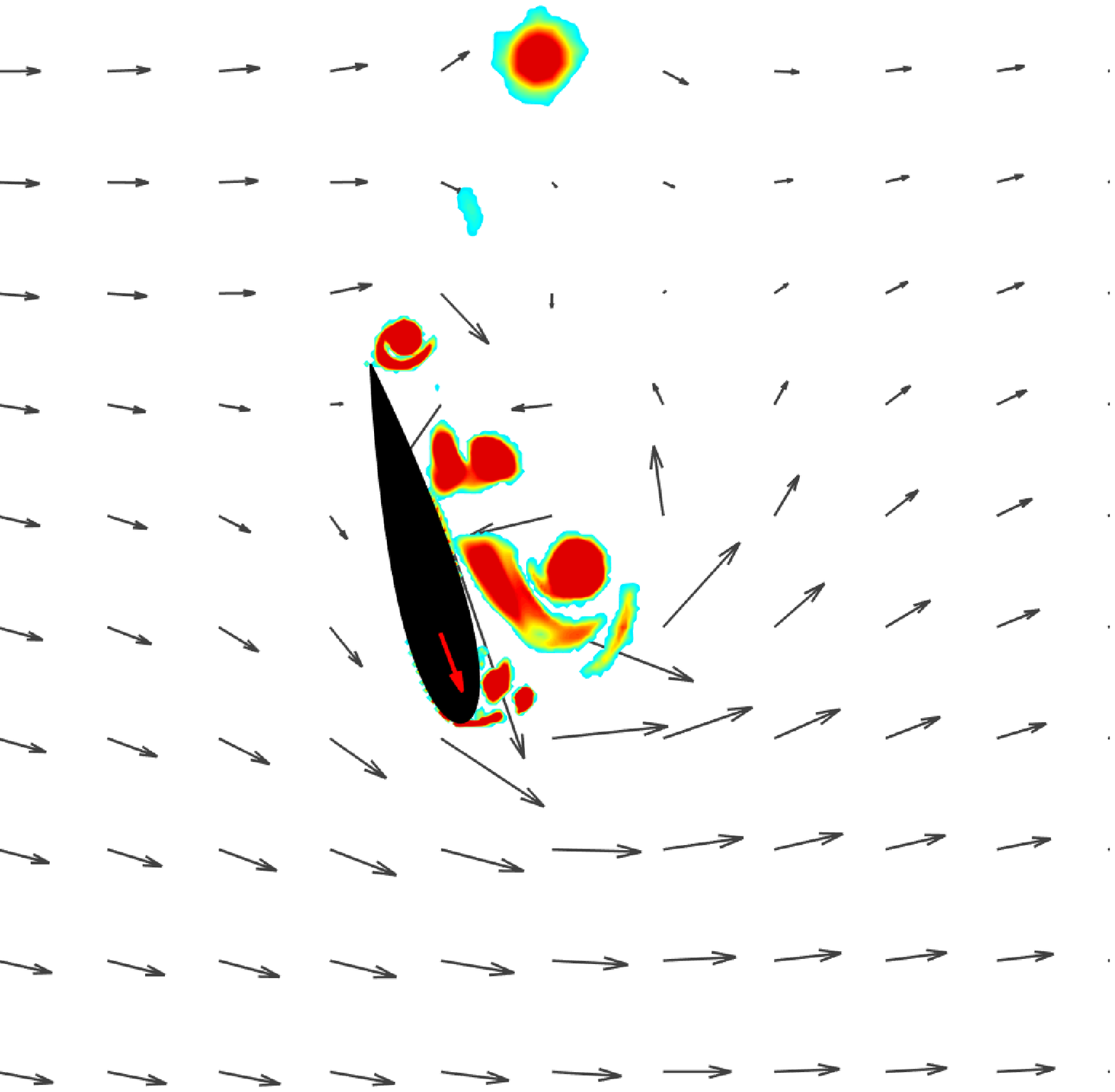}} \hfill
        \fbox{\includegraphics[height=1.25in]{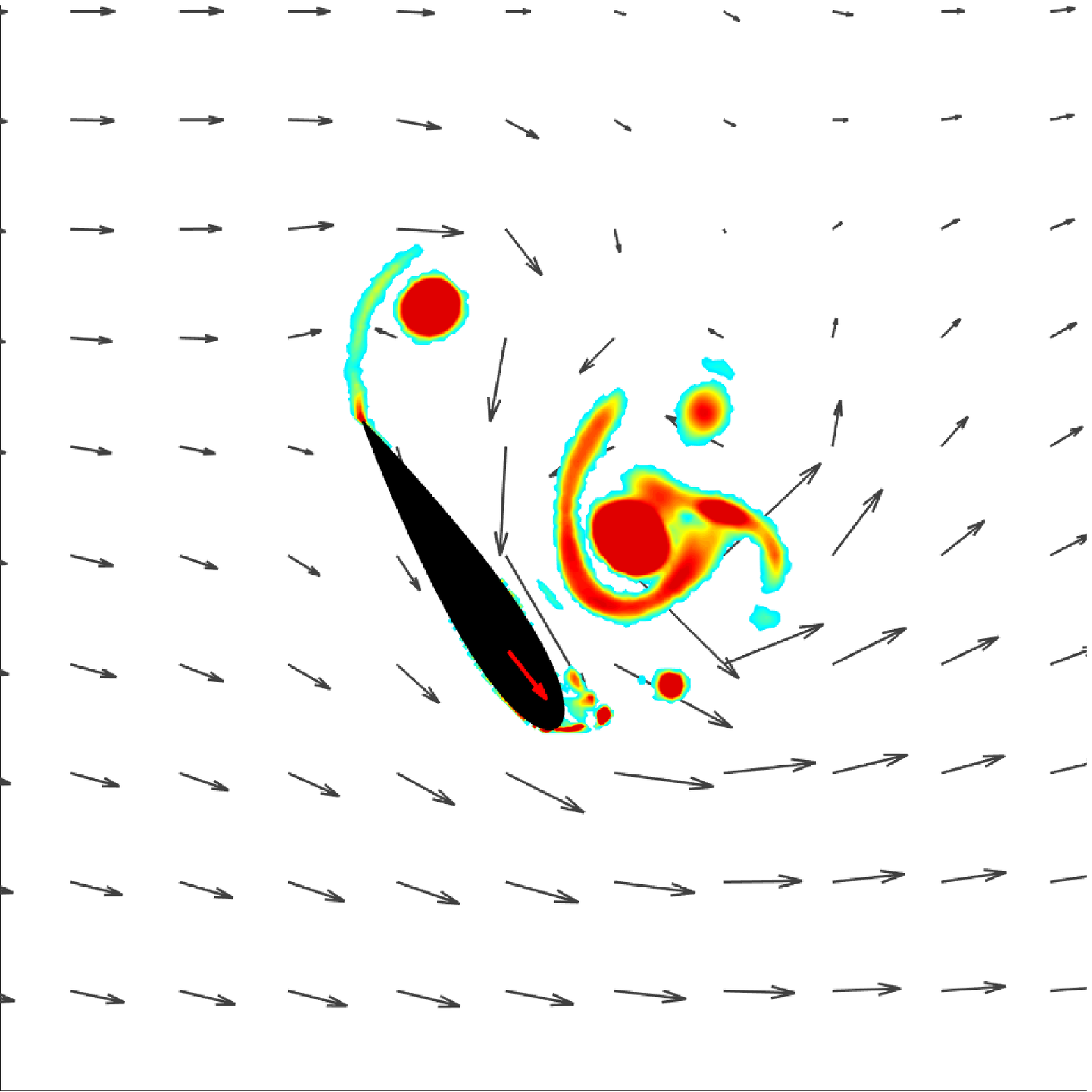}} \hfill
        \fbox{\includegraphics[height=1.25in]{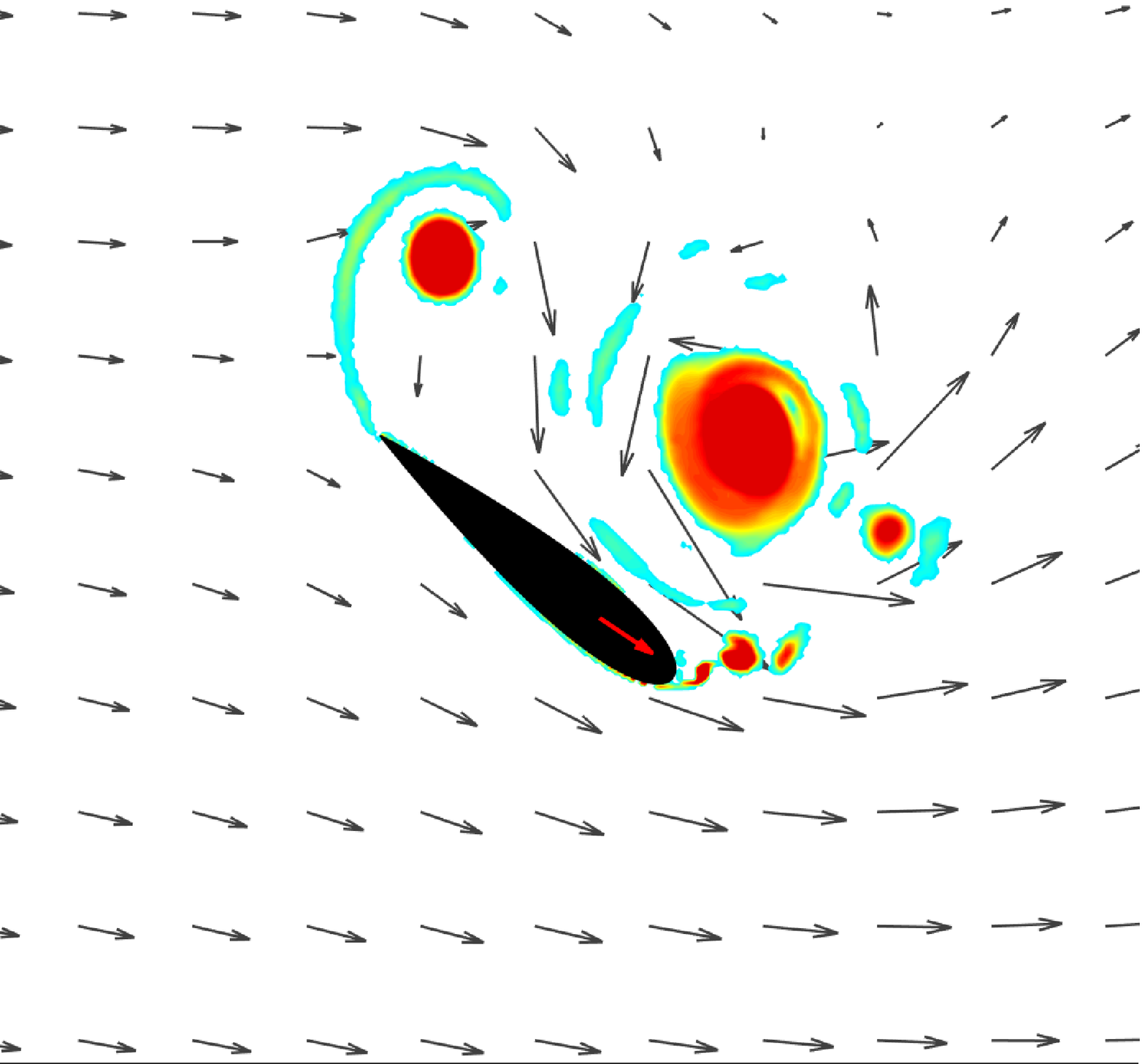}} \hfill
        \fbox{\includegraphics[height=1.25in]{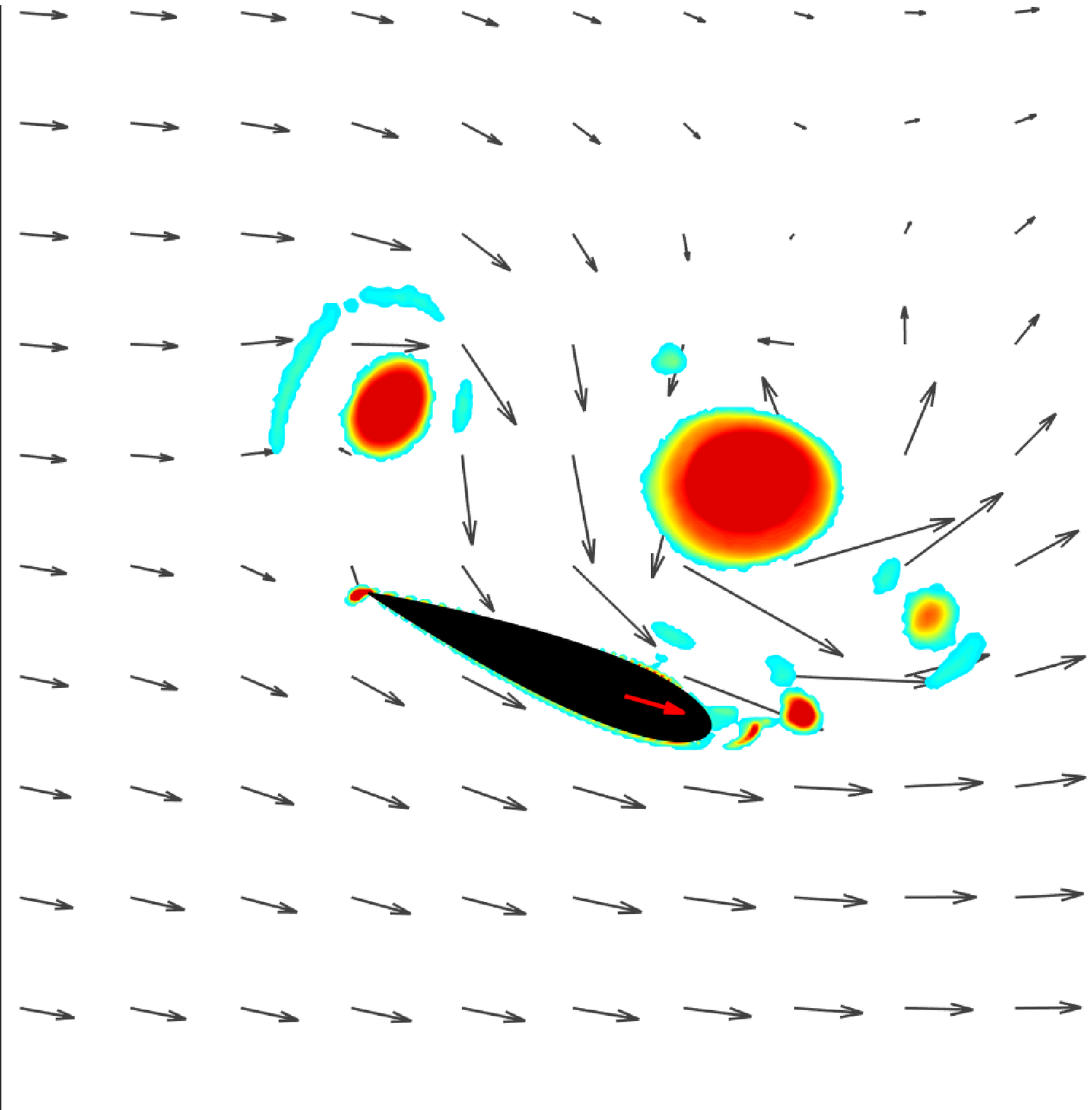}}
        \caption{URANS simulations, blockage ratio = 10.6\%}
        \label{subfig:rans_ff}
    \end{subfigure} \vspace{0.1in}
    
    \caption{(b)(c) Instantaneous flow velocity (vectors) and swirl strength (contours) from experimental data and URANS simulations show the generation of an LEV between $\theta=90^\circ$ to $\theta=180^\circ$ for $\lambda=1.1,\ Re = 4.5\times10^4$.}
    \label{fig:validation_flowfield}
\end{figure}

Results are also compared to experiments at a lower tip speed ratio. At $\lambda=1.1$, the blades experience higher angles of attack as seen in Fig. \ref{subfig:foil_kinematics_b}, and hence experience a deeper dynamic stall with separated flow over majority of the downstream sweep of the blade, which results in poor power generation. 
Figure \ref{subfig:1_1_val} compares the torque and forces for the confined flow computations with the experiments by Snortland et al. \cite{Snortland2019}. The peak power is over-predicted, along with discrepancies between $\theta=120^\circ$ to $\theta=180^\circ$.
Figures \ref{subfig:exp_ff} and \ref{subfig:rans_ff} directly compare the velocity and swirl strength from the computations with the phase-averaged particle image velocimetry (PIV) data from the experiments. The swirl strength is obtained by an eigenvalue analysis of the velocity gradient tensor \cite{ZHOU1999,Adrian2000}. It denotes the amount of local swirling motion of the flow and hence can be used to differentiate vortices from shear layers. The flow field shows the generation of an LEV between $\theta=90^\circ$ to $\theta=180^\circ$ due to flow separation on the upstream foil. The apparent angle of attack increases to its maximum value and then rapidly drops to zero during this part of the rotation (Fig. \ref{subfig:foil_kinematics_b}) which generates an LEV that is shed from the foil. This process corresponds to a drop in torque generation. The computed and experimental flow fields match well in terms of location and strength of the LEV, and the surrounding velocity field.

\red{The discrepancy in $C_P$ observed at $\lambda=1.1$ may be due to the difficulty of simulating highly separated flow using a URANS model. The two-dimensional approximation of the flow field cannot capture the three-dimensional instabilities inherent to high Reynolds number flows, including those within the boundary layer that affect transition and flow separation phenomena.}
Results also indicate high sensitivity to the Reynolds number in the regime explored, especially for the unconfined flow. \blue{The intrinsic transitional nature of the flow physics, and the varying of relative flow velocity throughout the cycle make this regime computationally challenging.} 
Increasing $Re$ from $4.5 \times 10^4$ to $1 \times 10^5$ for the unconfined configuration at $\lambda = 1.9$ increases the mean $C_P$ from 0.324 to 0.430, as shown in Table \ref{tab:meshes}. Due to the experimental conditions, the moderate Reynolds number of $4.5 \times 10^4$ remains the focus of this study.

\section{Intracycle Variation of Angular Velocity} \label{sec:results}

\subsection{Results for confined configuration}

\begin{figure}[t]
    \centering
    \begin{subfigure}{0.48\textwidth}
    \centering
    \includegraphics[width=\textwidth]{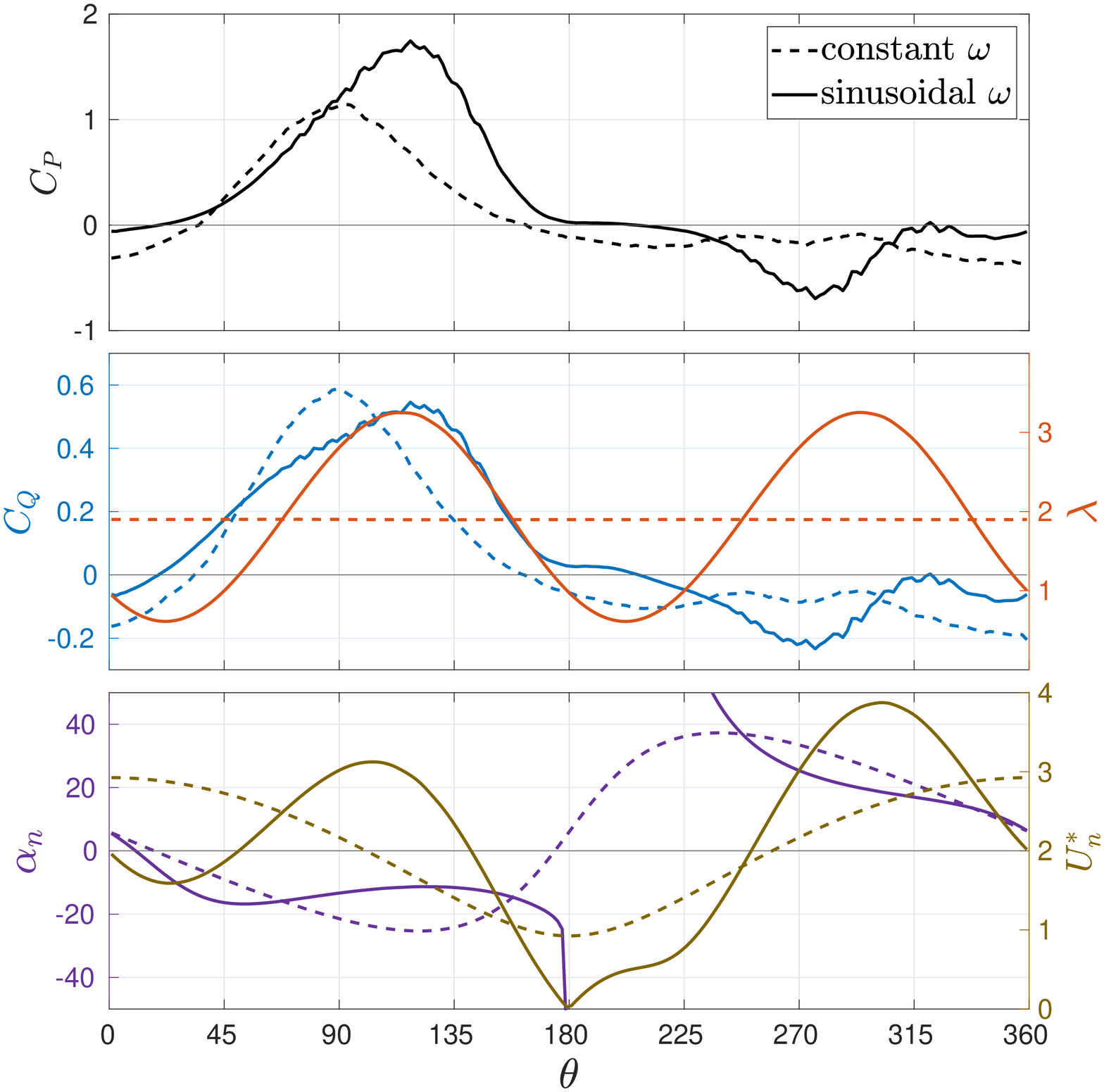}
    \caption{Experimental data, adapted from Strom et al. Blockage ratio = 11\%, $Re=3.1 \times 10^4$.}
    \label{subfig:strom}
    \end{subfigure} \hfill
    \begin{subfigure}{0.48\textwidth}
    \centering
    \includegraphics[width=\textwidth]{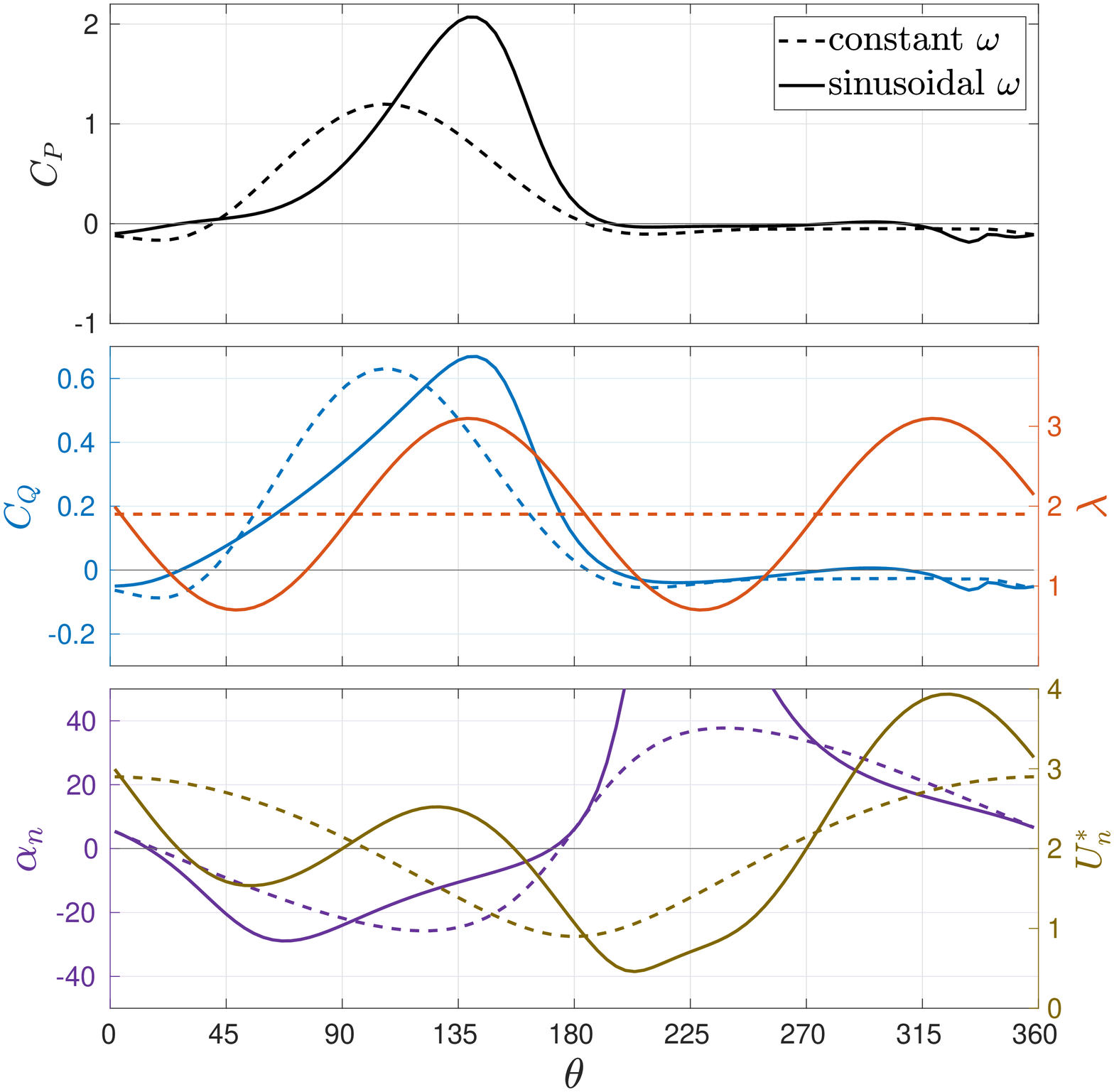}
    \caption{URANS computation results. Blockage ratio = 10.6\%, $Re=4.5 \times 10^4$.}
    \label{subfig:con}
    \end{subfigure}
    \caption{Comparison of performance for constant $\omega$ with that for sinusoidal variation of $\omega$ shows 53\% improvement in power generated for the experiments and 54.5\% for computations.}
    \label{fig:intracycle_performance}
\end{figure}

%% description of figure 7
A comparison of computational and experimental results for the confined flow configuration is presented in Fig. \ref{fig:intracycle_performance}.
Data for the baseline cases (constant $\omega$, $\lambda=1.9$) for the respective investigations are represented by dashed lines and those for the sinusoidal cases are represented by solid lines.
Figure \ref{subfig:strom} presents the kinematics and performance for the experiments, where the sinusoidal variation of $\omega$ (Eqn. \ref{eqn:omega_intra}) was optimized for maximum improvement in power coefficient compared to the baseline case, as discussed earlier in Strom et al. \cite{Strom2017}.
The experimental optimization was performed for a two blade turbine, but the data for a single blade turbine is presented to isolate the effect on each blade.
The variation of power coefficient, $C_P$ (black), is presented in the top frame, demonstrating the increased power generation with sinusoidal $\omega$. 
The measured phase-averaged torque coefficient, $C_Q$ (blue), is presented in the middle frame along with the tip speed ratio, $\lambda$ (orange), which is directly proportional to the angular velocity ($\lambda(\theta) = \omega(\theta) R / U_\infty$). The corresponding apparent angle of attack, $\alpha_n$ (purple), and normalized relative flow velocity, $U_n^*$ (green), calculated from Eqns. \ref{equation:alpha_n} and \ref{equation:u_nstar}, are shown in the bottom frame. 

Similarly, Fig. \ref{subfig:con} shows the computational data, where the dashed orange line in the middle frame shows the constant $\lambda=1.9$ for the baseline case and the solid orange line shows the sinusoidal variation.
The sinusoidal variation of angular velocity is prescribed as given by Eqn. \ref{eqn:omega_intra} with a mean value, $\overline{\omega}$, corresponding to $\overline{\lambda}=1.9$, and amplitude, $A_\omega=0.63\ \overline{\omega}$, very close to the optimized parameters in the experiment which has $A_\omega=0.68\ \overline{\omega}$.

There is a strong agreement between experiments and computations for the boost in power generation obtained with sinusoidal $\omega$, with an enhancement of 53\% in average $C_P$ for the experiments and 54.5\% for the computations.
The mechanism by which power generation is enhanced is that the peak of torque generation (blue lines) is delayed by approximately $30^\circ$ in terms of azimuthal position, which subsequently aligns with the peak angular velocity (orange lines). In order for these two peak values to align, the sinusoidal kinematics must be optimized by varying the phase with respect to azimuthal position, or $\phi_\omega$ in Eqn. \ref{eqn:omega_intra}. In the experimental data $\phi_{\omega}=\SI{3.96}{\radian}$ and in the computations $\phi_\omega = \SI{3.0}{\radian}$.
As the power generated is a product of torque and angular velocity, significantly more power is generated throughout the $\theta=110^\circ$ to $180^\circ$ portion of the cycle.

The exact location of maximum torque varies slightly between experiment and computation, which causes the above mentioned change in $\phi_\omega$. Directly comparing constant $\omega$ (dashed blue lines) in Fig. \ref{subfig:strom} and Fig. \ref{subfig:con}, the peak torque occurs at $\theta=90^\circ$ in the experiment and at $\theta=105^\circ$ in the simulation. Similarly for sinusoidal $\omega$ (solid blue lines), the peak torque is delayed until around $\theta=120^\circ$ for the experiment and until $\theta=140^\circ$ for the simulation.

It is important to note that it is the transient behavior of the intracycle kinematics that results in more power generation. During the intracycle control, the mean tip speed ratio is the same as the optimal tip speed ratio for a steady rotation rate. Thus, simply increasing $\omega$ throughout the rotation will result in a reduction of power generation. The variation in $\omega$ enables dramatic changes in the relative flow speed and angle of attack, modifying and likely \red{exploiting the dynamic stall process. In addition, when the sinusoidal variation is tuned properly, the maximum value of $\omega$ occurs at the same phase as the maximum $C_Q$ and thus, also enhances performance.
Because this is an outcome of the experimental optimization for this rotor geometry and inflow condition, this alignment might not be universal. Strom et al. \cite{Strom2017} includes a comparison between $C_P$ for constant speed and intracycle control, hypothesizing reasons for the differences throughout the cycle. Here, we explore the dynamic aspects of the flow field using the augmented data available through simulation.}

\begin{figure}[p]
    \centering
    \includegraphics[width=\textwidth]{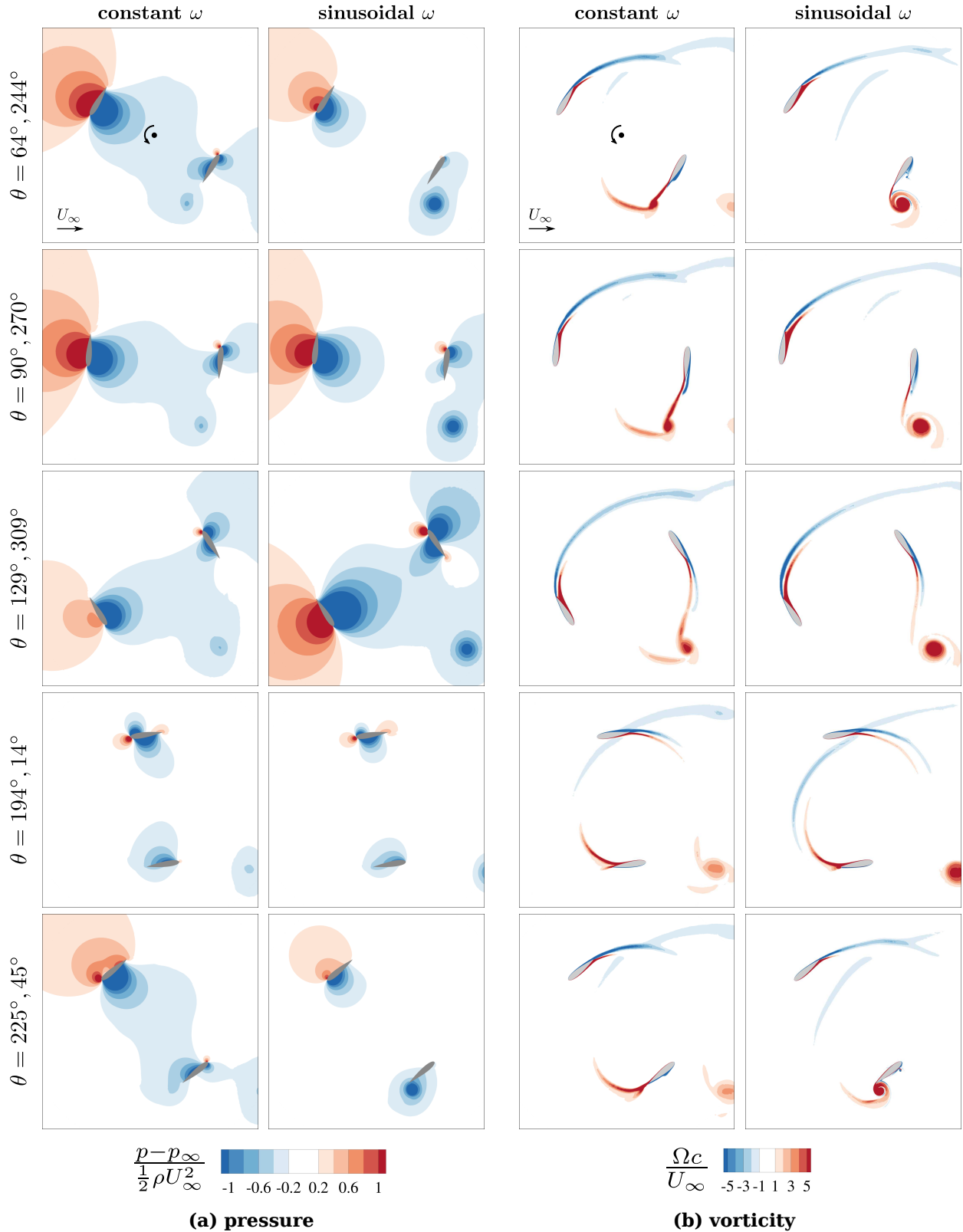}
    \caption{Normalized instantaneous pressure and vorticity fields for constant $\omega$ and for sinusoidal variation of $\omega$ from URANS simulations. Blockage ratio = 10.6\%, $Re=4.5 \times 10^4$.}
    \label{fig:con_fields}
\end{figure}

%%% FIg 8a and 8b
Figure \ref{fig:con_fields} displays the instantaneous pressure and vorticity fields at different azimuthal positions for the computations, directly contrasting the constant $\omega$ and sinusoidal $\omega$ flow fields. Examining the pressure contours of Fig. \ref{fig:con_fields}a demonstrates how the peak torque generation is delayed when compared to constant $\omega$.
At $\theta=64^\circ$, $\alpha_n$ is $15^\circ$ for constant $\omega$, whereas the sinusoidal $\omega$ has a higher $\alpha_n$ at $29^\circ$, but $U_n^*$ is relatively lower (Fig. \ref{subfig:con}). The result is a weaker pressure field on the blade for the sinusoidal $\omega$ in Fig. \ref{fig:con_fields}a, and as a consequence, less torque generation compared to constant $\omega$ for the this portion of the cycle.
%$C_P \approx 0.15$ as compared to $C_P \approx 0.55$ for constant $\omega$.
As the foil progresses to $\theta=90^\circ$, the relative flow velocity and angle of attack for both the cases are similar yet the torque generation for sinusoidal $\omega$ is still lower than that for constant $\omega$, and is increasing steadily with the increase in relative flow velocity. 
By the time the foil reaches $\theta=129^\circ$, the angle of attack for sinusoidal $\omega$ has decreased to around $12^\circ$, and the relative flow velocity is at its peak ($U_n^* \approx 2.5$). These changes can be seen in the strong pressure force in Fig. \ref{fig:con_fields}a as the torque generation exceeds that for constant $\omega$. This trend continues until the torque and blade velocity peak around $\theta=140^\circ$. On the other hand, the angle of attack for constant $\omega$ has risen to $\alpha_n=26^\circ$ and the relative velocity is much lower than that of the sinusoidal flow ($U_n^* \approx 1.5$), hence the torque generation begins decreasing after reaching its peak around $\theta=120^\circ$.

Throughout the downstream sweep of the turbine blade ($\theta=180^\circ$ to $\theta=360^\circ$), flow velocity encountered by the blade is low as a significant amount of energy has already been extracted from the flow during the upstream sweep of the blade. Hence the pressure field around the blade is substantially weaker than the that of the lead blade, and no significant torque is generated (Fig. \ref{subfig:con}). This is in contrast to the experiments where a significant amount of negative torque is observed (Fig. \ref{subfig:strom}) around $\theta=270^\circ$. The reason for this is unclear since flow field data is not yet available for the experiments with sinusoidal $\omega$.

\begin{figure}[p]
    \centering
    \includegraphics[width=0.8\textwidth]{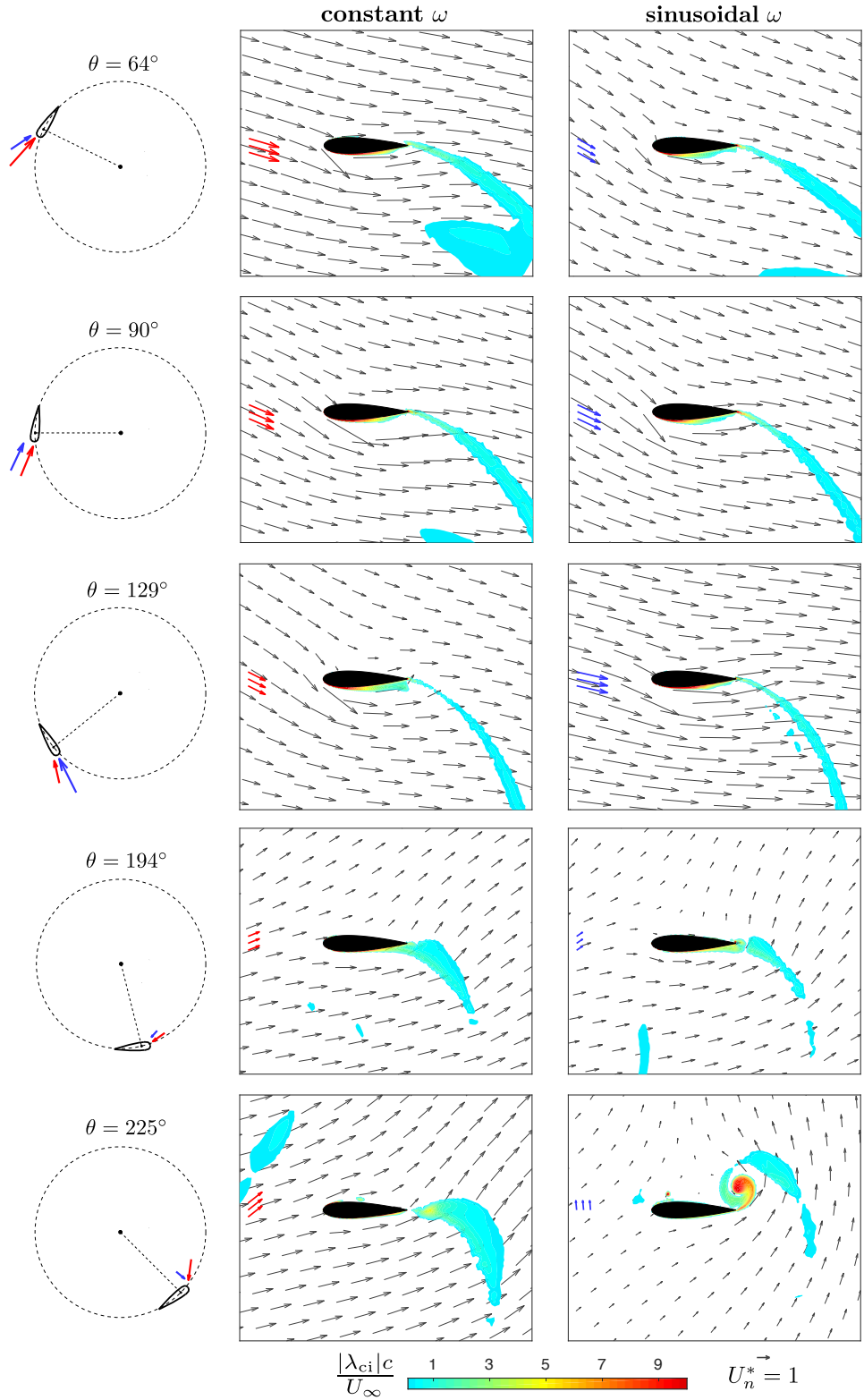}
    \caption{Flow velocity vectors in a rotating frame of reference \red{with the nominal relative flow velocity vectors for comparison; red (constant $\omega$) and blue (sinusoidal $\omega$).} Superimposed with contours of swirl strength. Blockage ratio = 10.6\%, $Re=4.5 \times 10^4$.}
    \label{fig:con_rel_velo}
\end{figure}

%% Fig 9
The vorticity fields in Fig. \ref{fig:con_fields}b show that there are no major vortices shed during the upstream phase of the blades for the confined flow configuration.
During the downstream phase, a single strong trailing edge vortex (TEV) is formed on the outward side of the blade around $\theta=225^\circ$ for the sinusoidal $\omega$. In contrast, for constant $\omega$ this TEV is relatively weak and develops fully downstream of the blade as opposed to on the blade itself.
The vortex generation can be analyzed in more detail through the local relative velocity of the flow. The \red{nominal} values of $\alpha_n$ and $U_n^*$ in Fig. \ref{fig:intracycle_performance} are calculated theoretically \red{(Eqns. \ref{equation:alpha_n} and \ref{equation:u_nstar})}, based on freestream velocity of oncoming flow, $U_\infty$. However these values can be significantly different from the actual flow velocities encountered by the blade due to induced flow effects.  \red{To explore this, Fig. \ref{fig:con_rel_velo} shows the velocity field from the simulation in a \textit{frame of reference rotating with the blade}. 
A strong indication of induction is apparent by comparing the nominal relative flow velocity vector (computed by Eqns. \ref{equation:alpha_n} and \ref{equation:u_nstar} in red (constant $\omega$) and blue (sinusoidal $\omega$)) with that of the actual flow field. The discrepancy is in both flow direction as well as the velocity magnitude, with reference to the unit vector in the bottom legend.
The nominal values of $\alpha_n$ and $U_n^*$ in this section are used to aid the visual representation of the flow field while also noting these differences due to flow induction.}
Superimposed with the flow vectors are contours of the swirl strength, which depicts the vortex formation on and around the blades.

{In the first row of Fig. \ref{fig:con_rel_velo}, $\alpha_n$ reaches its peak at $29^\circ$ for sinusoidal $\omega$ whereas for constant $\omega$, $\alpha_n$ is only $15^\circ$ and gradually increasing (Fig. \ref{subfig:con}). At this point in the cycle, $U_n^*$ for sinusoidal $\omega$ ($\approx 1.6$) is low compared to that for constant $\omega$ ($\approx 2.5$).
Hence for sinusoidal $\omega$, a separation bubble can be seen on the inward side (low pressure side) of the blade.
For constant $\omega$, the flow is mostly attached.
As the foil reaches $\theta=90^\circ$, $U_n^*$ and $\alpha_n$ for both cases are very similar as reflected in the similar flow fields and separation dynamics in Fig. \ref{fig:con_rel_velo}. Despite the similarities, the torque generation for sinusoidal $\omega$ is still lagging behind that of the constant $\omega$. The angle of attack for sinusoidal $\omega$ is decreasing and hence the flow starts reattaching to the inward side of the blade by $\theta=129^\circ$, however flow separation is observed for constant $\omega$ due to the high $\alpha_n$.}

After completing the upstream portion of the revolution, circulation around the blade is shed in the form of a trailing edge vortex (TEV).
$\alpha_n$ switches signs around $\theta=180^\circ$ and $U_n^*$ is low (Fig. \ref{subfig:con}) as the blade velocity is in the same direction as the freestream velocity. 
\red{This can be observed from the smaller velocity vectors on the outward side of the blade at $\theta=194^\circ$ in Fig. \ref{fig:con_rel_velo} (top of the blade).
However, the flow on the inward side (bottom of the blade in Fig. \ref{fig:con_rel_velo}) is highly decelerated due to energy extraction during the upstream phase, and hence has a higher velocity \textit{relative to the blade} observed by the longer velocity vectors. This sharp difference in relative flow} between the inward and outward side of the blade causes a roll-up of fluid around the trailing edge. % over a span of around $30^\circ$ to $50^\circ$. 
The circulation around the blade from its lift generation phase generates a vortex around $\theta=225^\circ$ (Fig. \ref{fig:con_fields}b).
These spatial and temporal changes in relative velocity are more drastic for the blade with sinusoidal angular velocity as seen in Fig. \ref{fig:con_rel_velo}. 
More circulation is shed in a short span as the torque generation drops from the peak to zero over a span of $40^\circ$ (Fig. \ref{subfig:con}), creating a relatively strong vortex (Fig. \ref{fig:con_fields}b). This TEV is shed from the blade as the blade velocity starts increasing again. The blade with constant $\omega$ on the other hand generates less lift during the upstream phase and hence less amount of circulation is released. The torque generation goes from the peak to zero over a longer span of $70^\circ$ resulting in a relatively weak vortex. For both the cases, the TEV is convected downstream and does not interact with the turbine blades any further (Fig. \ref{fig:con_fields}b).

\subsection{Results for unconfined configuration}

\begin{figure}[b!]
    \centering
    \includegraphics[width=0.48\textwidth]{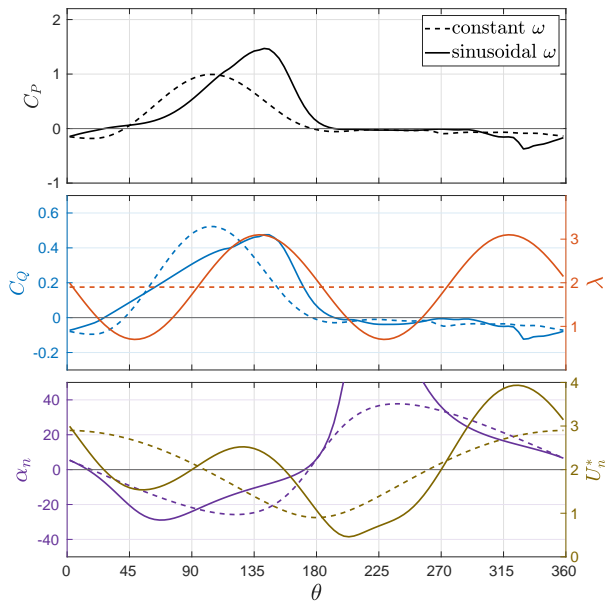}
    \caption{Comparison of URANS simulation results for constant $\omega$ with that for sinusoidal variation of $\omega$ show 27\% improvement in power generated for an unconfined configuration. Blockage ratio = 1.7\%, $Re=4.5 \times 10^4$.}
    \label{fig:intracycle_performance_open}
\end{figure}

Figure \ref{fig:intracycle_performance_open} shows the comparison between the torque and power generated from computations for a constant $\omega$ ($\lambda=1.9$) and sinusoidal $\omega$ for the unconfined configuration (low blockage ratio). The kinematic parameters of sinusoidal variation are the same as for the confined configuration. Using sinusoidal kinematics, an enhancement of 41\% is observed for the average power generation as compared to 54.5\% for the confined domain.

The peak of $C_Q$ (blue line) gets delayed by a similar azimuthal span as that for the confined configuration as it aligns with the peak of $\omega$ (orange line), but the peak value is lower than that reached in the confined flow. With sinusoidal $\omega$, the mean $C_P$ is 0.46 whereas for the confined configuration it is 0.68. This is consistent with the baseline constant $\omega$ in which the mean $C_P$ is lower than the confined configuration due to less blockage.

\begin{figure}[p]
    \centering
    \includegraphics[width=\textwidth]{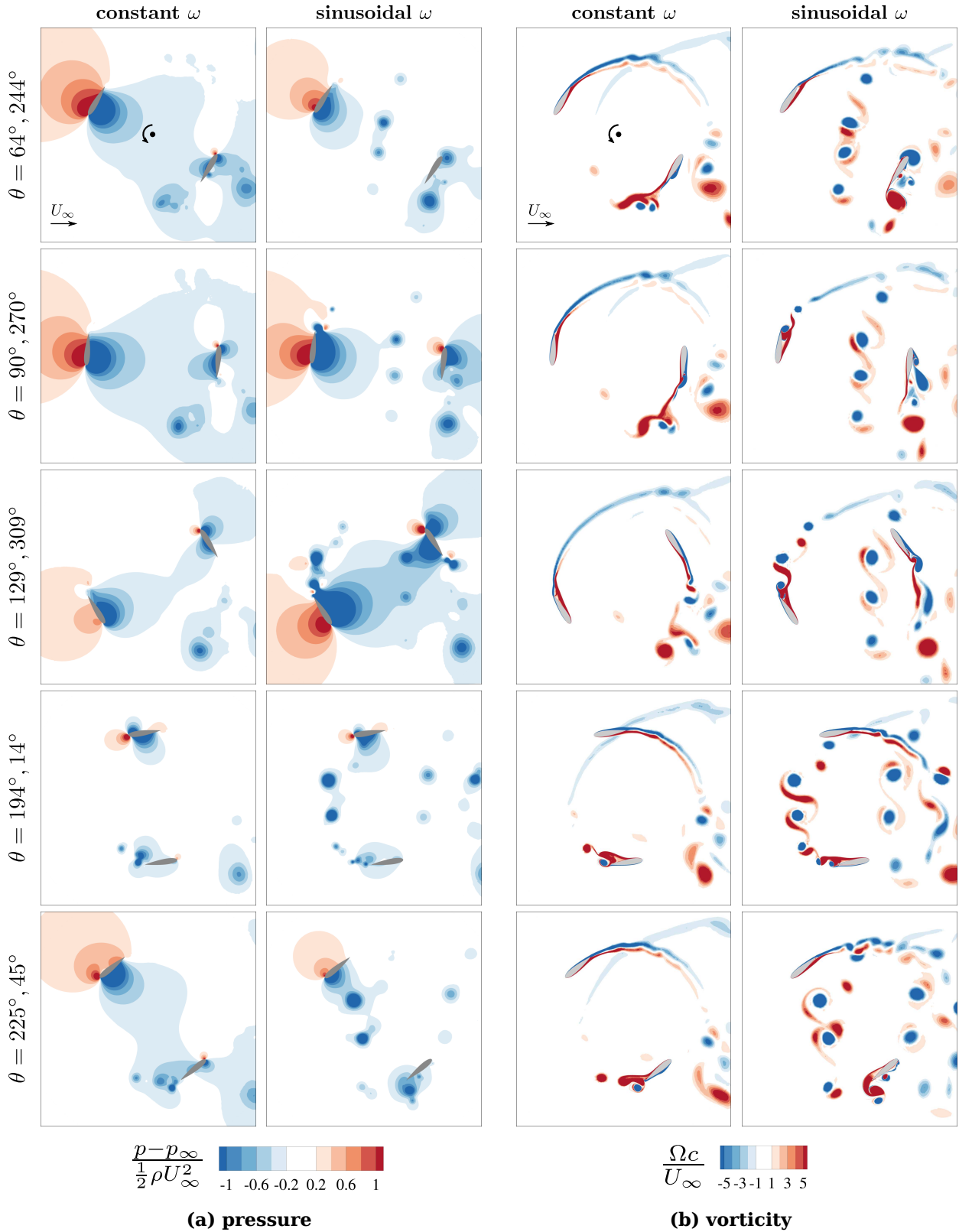}
    \caption{Normalized instantaneous pressure and vorticity fields for constant $\omega$ and for sinusoidal variation of $\omega$ from URANS simulations. Blockage ratio = 1.7\%, $Re=4.5 \times 10^4$.}
    \label{fig:open_fields}
\end{figure}

The pressure and vorticity fields are presented in Fig. \ref{fig:open_fields}. Contrasting Fig. \ref{fig:open_fields}a directly with Fig. \ref{fig:con_fields}a, the blockage effects are readily observed. 
\red{For incompressible flow, confinement results in higher flow velocity as the flow approaches the blade, and subsequently, sharper  pressure gradients around the blade. 
At lower confinement, the pressure gradients weaken since they extend over a larger region, as seen in Fig. \ref{fig:open_fields}a.}
The change in pressure distribution also affects the flow separation characteristics. For the confined configuration, the concentration of a low pressure region at the blade results in suppression or delay of flow separation and thus the blade sustains high torque generation. Even as flow separates at high angles of attack, a separation bubble is formed but the flow quickly reattaches and does not shed leading edge vortices.
On the other hand, the unconfined configuration has relatively lower forces on the blade, which is reflected in the torque generation. The pressure field is likely causing a less favorable pressure gradient, leaving the boundary layer more susceptible to flow separation.   
The separation negatively affects the torque generation since it reduces the component of force in the direction of motion of the blade. Hence as observed in Fig. \ref{fig:intracycle_performance_open}, $C_Q$ does not increase significantly after $\theta=90^\circ$ for sinusoidal $\omega$ although it reaches its peak at a similar azimuthal position as confined flow. 

\begin{figure}[p]
    \centering
    \includegraphics[width=0.8\textwidth]{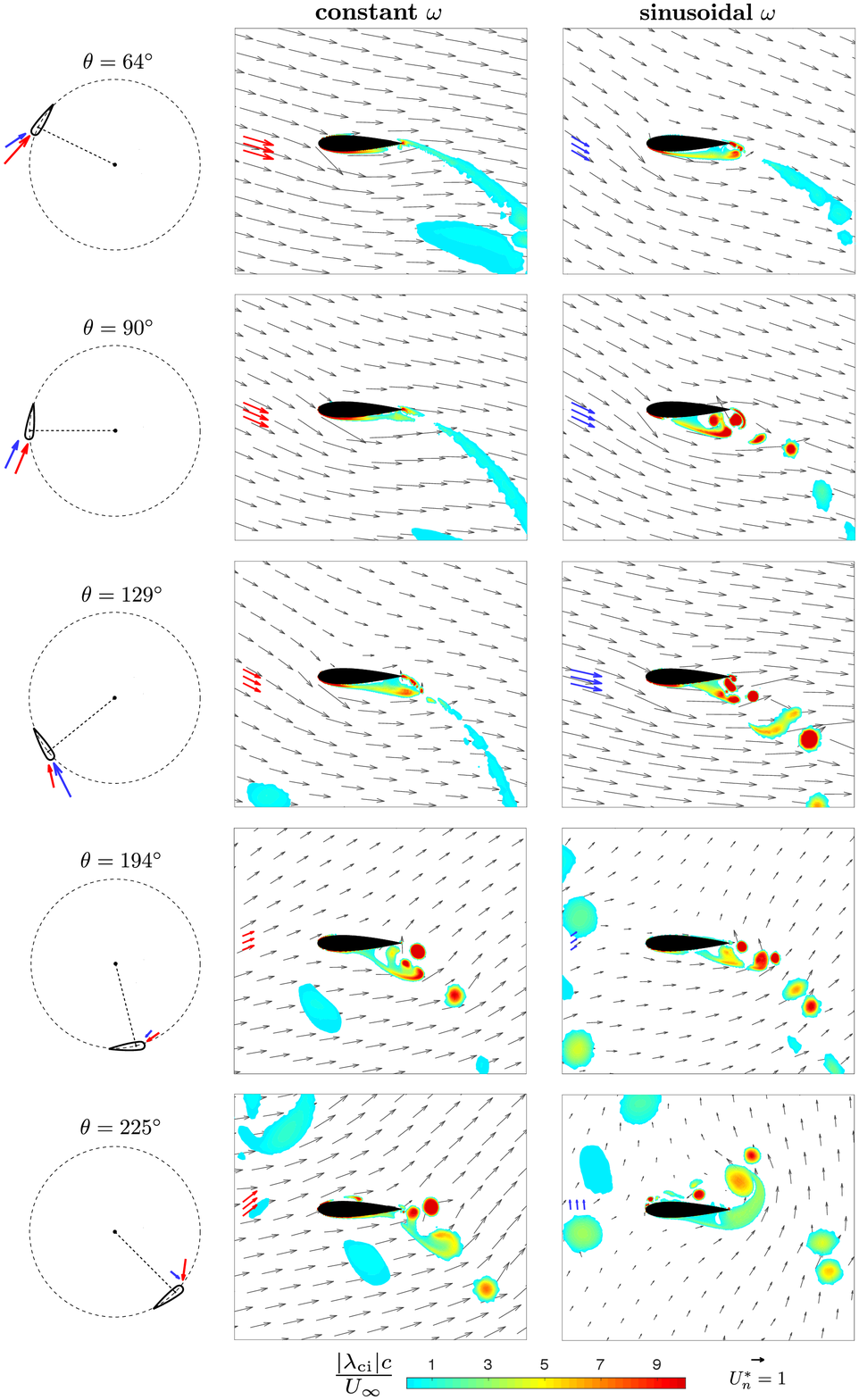}
    \caption{Flow velocity vectors in a rotating frame of reference \red{with the nominal relative flow velocity vectors for comparison; red (constant $\omega$) and blue (sinusoidal $\omega$).} Superimposed with contours of swirl strength. Blockage ratio = 1.7\%, $Re=4.5 \times 10^4$.}
    \label{fig:uncon_rel_velo}
\end{figure}

\red{Coupled to the flow separation is the formation of} multiple vortices that are generated at the blade for sinusoidal $\omega$ during its upstream sweep, as shown in Figure \ref{fig:open_fields}b. These vortices do interact with the second blade during its downstream sweep in the next rotation of the turbine, but due to low velocities there is no notable effect of the blade-vortex interaction on $C_Q$ and $C_P$ (Fig. \ref{fig:intracycle_performance_open}). \red{While an analysis of detailed flow separation dynamics leading to the generation of multiple vortices is not included in this investigation due to the limitations of two-dimensional URANS modeling explained in section \ref{sec:mesh}, the effect of intracycle variation of $\omega$ on the surrounding flow field and torque generation is investigated. Along with the pressure field in Fig. \ref{fig:open_fields}a, this is explained} through velocities in reference frame of the blade along with the swirl strength, as presented in Fig. \ref{fig:uncon_rel_velo}.
For sinusoidal $\omega$ at $\theta=64^\circ$, when the angle of attack is at its highest ($\alpha = 29^\circ$), the flow field shows a much larger flow separation (Fig. \ref{fig:uncon_rel_velo}) than the confined configuration (Fig. \ref{fig:con_rel_velo}), as the boundary layer separates from the leading edge.
This subsequently forms a recirculation region shown at $\theta=90^\circ$ in which the boundary layer rolls up into a coherent leading edge vortex (LEV) by the trailing edge. The vortex is shed from the surface with a counter-rotating TEV also depicted in Fig. \ref{fig:open_fields}b. \red{Fig. \ref{fig:open_fields}a also shows a stronger low pressure region by the trailing edge due to these vortices that reduces the component of force exerted in the direction of blade velocity.} Further along in the cycle, two more vortex pairs are shed as the angle of attack is decreasing and the flow reattaches to the blade (Fig. \ref{fig:uncon_rel_velo}). 
However the counter-clockwise LEVs are deformed under the influence of the clockwise TEVs and dissipate more rapidly.

Following the sinusoidal kinematics for the downstream sweep, a TEV is generated from $\theta=180^\circ$ to $\theta=225^\circ$ after the angle of attack switches sides, releasing the circulation around the blade from its lift generation phase during the upstream sweep. The TEV sheds approximately $10^\circ$ earlier for the unconfined configuration as compared to the confined configuration due to more flow separation, and it gets convected downstream without interacting any further with the blade. 

Examining the relative velocity in Fig. \ref{fig:uncon_rel_velo} for constant $\omega$, the angle of attack increases gradually and reaches its peak ($\alpha_n=26^\circ$) at $\theta=129^\circ$. At this large angle of attack, the flow separates but there is no vortex shedding. After $\theta=129^\circ$ the angle of attack decreases rapidly, and a streak of vorticity is shed along with a pair of counter-rotating vortices. This phenomenon is similar to the confined flow, but begins around $\theta=180^\circ$, earlier than confined flow where the vorticity is shed around $\theta=225^\circ$. The shed vorticity rolls into a vortex, is convected downstream, and is stronger than that of the confined flow due to a higher degree of flow separation. Since there are no vortices generated during the upstream sweep, there are no vortex-blade interactions.

\subsection{Dependence on flow conditions} 

The results above demonstrated that there is a difference in vortex generation and pressure fields between the confined and unconfined conditions tested.  The confined flow configuration was designed to match previous flume experiments, however the unconfined configuration is investigated to rule out phenomena that could be purely associated with blockage.  

As predicted, more blockage creates higher forces, more torque, and faster fluid acceleration which suppresses the flow separation. The unconfined flow configuration sees poorer performance than the confined flow in both the constant and sinusoidal kinematics. However, both configurations saw a significant increase in power, 41\% for the unconfined flow, and 54.5\% for the confined flow.

However it should be noted that the kinematics were identical for both confined and unconfined configurations. It could be that the optimal parameters for sinusoidal variation of the turbine angular velocity in an unconfined configuration vary from those for the confined configuration. Likewise, $\lambda=1.9$ may not be the optimum tip speed ratio for the specified turbine in the unconfined flow configuration. Hence, a separate optimization study for the unconfined configuration with a constant and intracycle variation of angular velocity may yield higher performance enhancements. 

It has also been shown in the baseline flow (constant $\omega$) in section \ref{sec:mesh} that the unconfined flow configuration is sensitive to Reynolds number changes.  \blue{This is likely explained by the prominent flow separation shown in Fig. \ref{fig:open_fields}, which can be further suppressed at higher Reynolds number.} \red{Less severe boundary layer separation correlates with higher torque generation. Furthermore, the sensitivity to $Re$ (and thus flow separation) could partially be a consequence of the limitations of two-dimensional URANS modeling in this transitional regime as discussed in section \ref{sec:mesh}.} The baseline flow in Table \ref{tab:meshes} demonstrates a 32.7\% increase in power coefficient when the Reynolds number is increased from $Re=4.5 \times 10^4$ to $Re=1 \times 10^5$. In this regime there is much less separation and the flow fields more closely resemble that of the lower Reynolds number confined flow.

\begin{figure}[t]
    \centering
    \begin{subfigure}{0.48\textwidth}
        \centering
        \includegraphics[width=\textwidth]{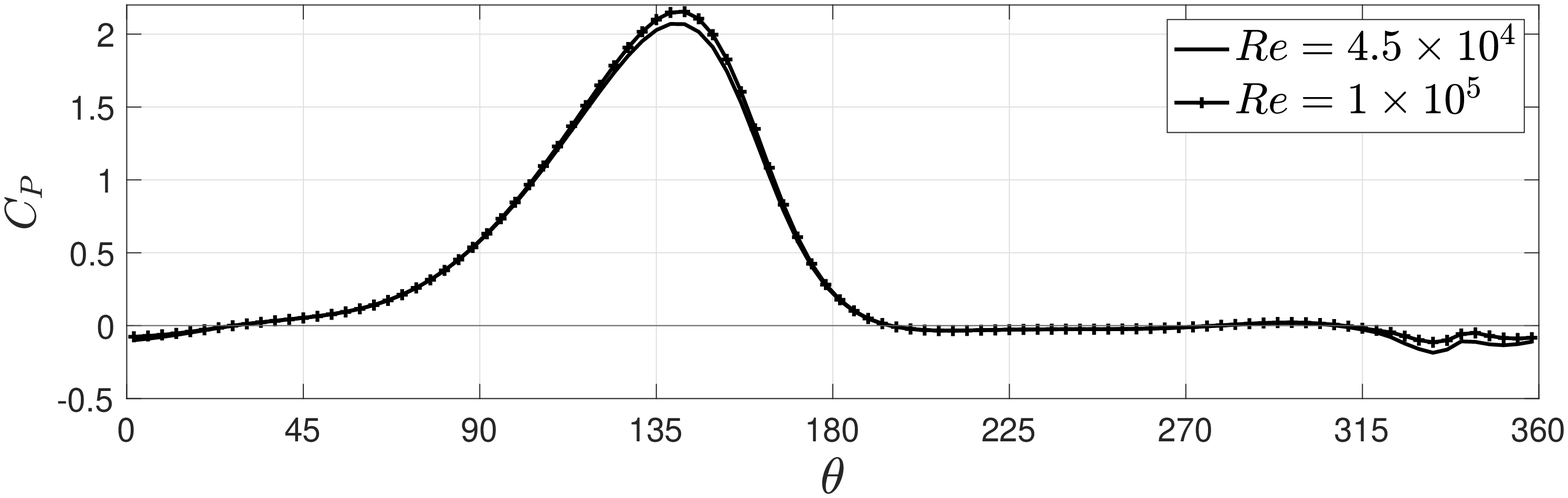}
        \caption{Confined configuration}
        \label{subfig:Re_con}
    \end{subfigure}
    \hfill
    \begin{subfigure}{0.48\textwidth}
        \centering
        \includegraphics[width=\textwidth]{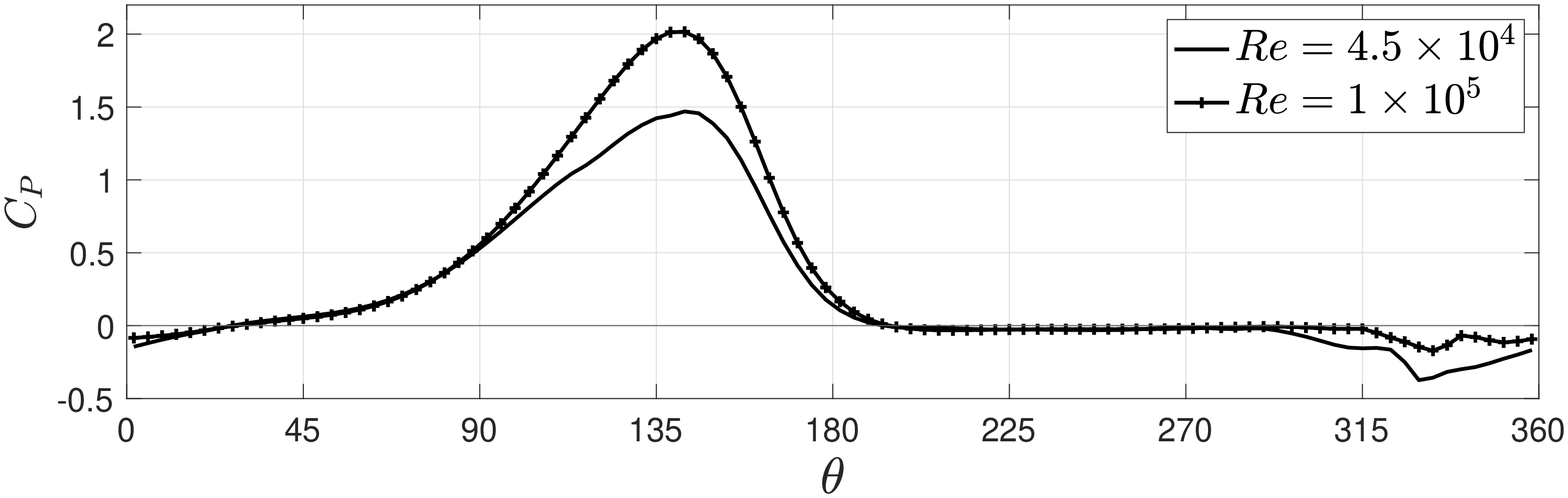}
        \caption{Unconfined configuration}
        \label{subfig:Re_uncon}
    \end{subfigure}
    \caption{Comparison of the power coefficient with sinusoidal variation of $\omega$, demonstrating high Reynolds number dependence for the unconfined configuration.}
    \label{fig:Re_comp}
\end{figure}

Likewise, when implementing the sinusoidal kinematics at $Re=1 \times 10^5$, the improvement in efficiency from a constant $\omega$ jumps to 53.5\% for the unconfined configuration. This boost in power coefficient is roughly proportional to the 54.5\% improvement shown for the confined flow at $Re=4.5 \times 10^4$. The Reynolds number dependence for sinusoidal kinematics is shown in Fig. \ref{fig:Re_comp}.

\section{Conclusion}

A new strategy for improving the efficiency of straight-bladed cross-flow turbines is explored by intracycle variation of its angular velocity. Both a confined configuration with blockage ratio of 10.6\% and an unconfined configuration with a low blockage ratio of 1.7\% are investigated computationally using a 2D URANS model.

For the confined configuration, a significant average power enhancement of 54.5\% is observed for sinusoidally varying angular velocity control compared to a constant rotation rate scheme at the same mean tip speed ratio. 
The overall power enhancement is very close to the 53\% enhancement achieved by Strom et al. \cite{Strom2017}, and these simulations provide valuable time-resolved flow fields of the on-blade separation and subsequent vortex dynamics. 
The time-evolution of the flow relative to the blade plays a critical role in the power enhancement, as the angle of attack increases rapidly at the start of the cycle and then decreases gradually while the relative velocity rises and the peak torque generation aligns with the peak of angular velocity. Throughout this process, there is little flow separation except a single, strong trailing edge vortex that is formed after the end of the upstream sweep of the blade.

For the unconfined configuration, the sinusoidal variation in angular velocity increases the average power generation by 41\% compared to constant angular velocity control. The lack of confinement results in more flow separation and multiple shed vortices that interact with the blades during their downstream sweep. The intracycle kinematics are the same as for the confined case, however due to the difference in pressure gradients associated with confinement, a separate optimization process may reveal a different set of optimal kinematics for the unconfined configuration.
All simulations are performed at a Reynolds number (based on chord and freestream) of $4.5\times10^4$ to match those of the constant angular velocity experiments for validation purposes. Further simulations were also conducted at a higher Reynolds number of $1\times10^5$ for the unconfined case. In this regime the power increased significantly to a 53.5\% improvement, more or less the same as the confined flow.

\blue{URANS computations such as those performed here often have limitations when modeling flows with large separation, in addition to those at moderate Reynolds numbers within a transitional regime. The computations, both constant and sinusoidal angular velocity, compare well with experiments using the current model. However more insight into the flow physics may be obtained with three-dimensional URANS or higher-fidelity large-eddy simulations (LES).}

\red{In terms of power generation, our findings depend on factors such as the Reynolds number and the turbine geometry (i.e., the number and cross-section of turbine blades, radius of the turbine, and the preset pitch angle).} However, this investigation provides insight into the unsteady flow dynamics and processes by which the mean power is enhanced via sinusoidal, intracycle, angular velocity in a cross-flow turbine. A thorough understanding of these dynamic mechanisms will lead to better modeling and optimization tools for cross-flow turbines under a variety of non-uniform or array configurations.

\section*{Acknowledgments}
This research was conducted using computational resources at the Center for Computation and Visualization, Brown University. University of Washington researchers were supported by the U.S. Department of Defense Naval Facilities Engineering Command (contract no. N0002418F8702). Abigale Snortland would like to acknowledge the National Physical Science Consortium for a one year fellowship.

\bibliography{darrieus,misc_ref}

\begin{thebibliography}{56}
\newcommand{\enquote}[1]{``#1''}
\providecommand{\natexlab}[1]{#1}
\providecommand{\url}[1]{\texttt{#1}}
\providecommand{\urlprefix}{URL }
\expandafter\ifx\csname urlstyle\endcsname\relax
  \providecommand{\doi}[1]{\discretionary{}{}{}https://doi.org/#1}\else
  \providecommand{\doi}[1]{\discretionary{}{}{}\urlstyle{rm}\url{https://doi.org/#1}}\fi

\bibitem[{Kirke and Lazauskas(2011)}]{Kirke2011}
Kirke, B.~K., and Lazauskas, L., \enquote{{Limitations of fixed pitch Darrieus
  hydrokinetic turbines and the challenge of variable pitch},} \emph{Renew.
  Energy}, Vol.~36, No.~3, 2011, pp. 893--897.
\newblock \doi{10.1016/j.renene.2010.08.027}.

\bibitem[{{Aslam Bhutta} et~al.(2012){Aslam Bhutta}, Hayat, Farooq, Ali, Jamil,
  and Hussain}]{AslamBhutta2012}
{Aslam Bhutta}, M.~M., Hayat, N., Farooq, A.~U., Ali, Z., Jamil, S.~R., and
  Hussain, Z., \enquote{{Vertical axis wind turbine - A review of various
  configurations and design techniques},} \emph{Renew. Sustain. Energy Rev.},
  Vol.~16, No.~4, 2012, pp. 1926--1939.
\newblock \doi{10.1016/j.rser.2011.12.004}.

\bibitem[{Timmer(2008)}]{Timmer2008}
Timmer, W., \enquote{{Two-Dimensional Low-Reynolds Number Wind Tunnel Results
  for Airfoil NACA 0018},} \emph{Wind Eng.}, Vol.~32, No.~6, 2008, pp.
  525--537.
\newblock \doi{10.1260/030952408787548848}.

\bibitem[{Fujisawa and Shibuya(2001)}]{Fujisawa2001}
Fujisawa, N., and Shibuya, S., \enquote{{Observations of dynamic stall on
  turbine blades},} \emph{J. Wind Eng. Ind. Aerodyn.}, Vol.~89, No.~2, 2001,
  pp. 201--214.
\newblock \doi{10.1016/S0167-6105(00)00062-3}.

\bibitem[{{Sim{\~{a}}o Ferreira} et~al.(2009){Sim{\~{a}}o Ferreira}, van Kuik,
  van Bussel, and Scarano}]{SimaoFerreira2009}
{Sim{\~{a}}o Ferreira}, C., van Kuik, G., van Bussel, G., and Scarano, F.,
  \enquote{{Visualization by PIV of dynamic stall on a vertical axis wind
  turbine},} \emph{Exp. Fluids}, Vol.~46, No.~1, 2009, pp. 97--108.
\newblock \doi{10.1007/s00348-008-0543-z}.

\bibitem[{Dunne and McKeon(2015)}]{Dunne2015}
Dunne, R., and McKeon, B.~J., \enquote{{Dynamic stall on a pitching and surging
  airfoil},} \emph{Exp. Fluids}, Vol.~56, No.~8, 2015, p. 157.
\newblock \doi{10.1007/s00348-015-2028-1}.

\bibitem[{Buchner et~al.(2015)Buchner, Lohry, Martinelli, Soria, and
  Smits}]{Buchner2015}
Buchner, A.~J., Lohry, M.~W., Martinelli, L., Soria, J., and Smits, A.~J.,
  \enquote{{Dynamic stall in vertical axis wind turbines: Comparing experiments
  and computations},} \emph{J. Wind Eng. Ind. Aerodyn.}, Vol. 146, 2015, pp.
  163--171.
\newblock \doi{10.1016/j.jweia.2015.09.001}.

\bibitem[{{Sim{\~{a}}o Ferreira} et~al.(2010){Sim{\~{a}}o Ferreira}, van
  Zuijlen, Bijl, van Bussel, and van Kuik}]{SimaoFerreira2010}
{Sim{\~{a}}o Ferreira}, C.~J., van Zuijlen, A., Bijl, H., van Bussel, G., and
  van Kuik, G., \enquote{{Simulating dynamic stall in a two-dimensional
  vertical-axis wind turbine: verification and validation with particle image
  velocimetry data},} \emph{Wind Energy}, Vol.~13, No.~1, 2010, pp. 1--17.
\newblock \doi{10.1002/we.330}.

\bibitem[{Wang et~al.(2010)Wang, Ingham, Ma, Pourkashanian, and Tao}]{Wang2010}
Wang, S., Ingham, D.~B., Ma, L., Pourkashanian, M., and Tao, Z.,
  \enquote{{Numerical investigations on dynamic stall of low Reynolds number
  flow around oscillating airfoils},} \emph{Comput. Fluids}, Vol.~39, No.~9,
  2010, pp. 1529--1541.
\newblock \doi{10.1016/j.compfluid.2010.05.004}.

\bibitem[{Tsai and Colonius(2016)}]{Tsai2016}
Tsai, H.-C., and Colonius, T., \enquote{{Coriolis Effect on Dynamic Stall in a
  Vertical Axis Wind Turbine},} \emph{AIAA J.}, Vol.~54, No.~1, 2016, pp.
  216--226.
\newblock \doi{10.2514/1.J054199}.

\bibitem[{Wang et~al.(2016)Wang, Sun, Zhu, Zhang, and Huang}]{Wang2016}
Wang, Y., Sun, X., Zhu, B., Zhang, H., and Huang, D., \enquote{{Effect of blade
  vortex interaction on performance of Darrieus-type cross flow marine current
  turbine},} \emph{Renew. Energy}, Vol.~86, 2016, pp. 316--323.
\newblock \doi{10.1016/j.renene.2015.07.089}.

\bibitem[{Almohammadi et~al.(2015)Almohammadi, Ingham, Ma, and
  Pourkashanian}]{Almohammadi2015}
Almohammadi, K., Ingham, D., Ma, L., and Pourkashanian, M., \enquote{{Modeling
  dynamic stall of a straight blade vertical axis wind turbine},} \emph{J.
  Fluids Struct.}, Vol.~57, 2015, pp. 144--158.
\newblock \doi{10.1016/j.jfluidstructs.2015.06.003}.

\bibitem[{Lei et~al.(2017)Lei, Zhou, Bao, Li, and Han}]{Lei2017}
Lei, H., Zhou, D., Bao, Y., Li, Y., and Han, Z., \enquote{{Three-dimensional
  Improved Delayed Detached Eddy Simulation of a two-bladed vertical axis wind
  turbine},} \emph{Energy Convers. Manag.}, Vol. 133, 2017, pp. 235--248.
\newblock \doi{10.1016/j.enconman.2016.11.067}.

\bibitem[{Hand et~al.(2017)Hand, Kelly, and Cashman}]{Hand2017}
Hand, B., Kelly, G., and Cashman, A., \enquote{{Numerical simulation of a
  vertical axis wind turbine airfoil experiencing dynamic stall at high
  Reynolds numbers},} \emph{Comput. Fluids}, Vol. 149, 2017, pp. 12--30.
\newblock \doi{10.1016/j.compfluid.2017.02.021}.

\bibitem[{L{\'{o}}pez et~al.(2016)L{\'{o}}pez, Meneses, Quintero, and
  La{\'{i}}n}]{Lopez2016}
L{\'{o}}pez, O., Meneses, D., Quintero, B., and La{\'{i}}n, S.,
  \enquote{{Computational study of transient flow around Darrieus type cross
  flow water turbines},} \emph{J. Renew. Sustain. Energy}, Vol.~8, No.~1, 2016,
  p. 014501.
\newblock \doi{10.1063/1.4940023}.

\bibitem[{Bachant and Wosnik(2016)}]{Bachant2016}
Bachant, P., and Wosnik, M., \enquote{{Modeling the near-wake of a
  vertical-axis cross-flow turbine with 2-D and 3-D RANS},} \emph{J. Renew.
  Sustain. Energy}, Vol.~8, No.~5, 2016, p. 053311.
\newblock \doi{10.1063/1.4966161}.

\bibitem[{Ouro et~al.(2018)Ouro, Stoesser, and Ram{\'{i}}rez}]{Ouro2018}
Ouro, P., Stoesser, T., and Ram{\'{i}}rez, L., \enquote{{Effect of Blade
  Cambering on Dynamic Stall in View of Designing Vertical Axis Turbines},}
  \emph{J. Fluids Eng.}, Vol. 140, No.~6, 2018.
\newblock \doi{10.1115/1.4039235}.

\bibitem[{Amet et~al.(2009)Amet, Ma{\^{i}}tre, Pellone, and Achard}]{Amet2009}
Amet, E., Ma{\^{i}}tre, T., Pellone, C., and Achard, J.~L., \enquote{{2D
  numerical simulations of blade-vortex interaction in a darrieus turbine},}
  \emph{J. Fluids Eng. Trans. ASME}, Vol. 131, No.~11, 2009, pp. 1--15.
\newblock \doi{10.1115/1.4000258}.

\bibitem[{Wang and Zhuang(2017)}]{Wang2017}
Wang, Z., and Zhuang, M., \enquote{{Leading-edge serrations for performance
  improvement on a vertical-axis wind turbine at low tip-speed-ratios},}
  \emph{Appl. Energy}, Vol. 208, 2017, pp. 1184--1197.
\newblock \doi{10.1016/j.apenergy.2017.09.034}.

\bibitem[{Sobhani et~al.(2017)Sobhani, Ghaffari, and Maghrebi}]{Sobhani2017}
Sobhani, E., Ghaffari, M., and Maghrebi, M.~J., \enquote{{Numerical
  investigation of dimple effects on darrieus vertical axis wind turbine},}
  \emph{Energy}, Vol. 133, 2017, pp. 231--241.
\newblock \doi{10.1016/j.energy.2017.05.105}.

\bibitem[{Xiao et~al.(2013)Xiao, Liu, and Incecik}]{Xiao2013}
Xiao, Q., Liu, W., and Incecik, A., \enquote{{Flow control for VATT by fixed
  and oscillating flap},} \emph{Renew. Energy}, Vol.~51, 2013, pp. 141--152.
\newblock \doi{10.1016/j.renene.2012.09.021}.

\bibitem[{Yen and Ahmed(2013)}]{Yen2013}
Yen, J., and Ahmed, N.~A., \enquote{{Enhancing vertical axis wind turbine by
  dynamic stall control using synthetic jets},} \emph{J. Wind Eng. Ind.
  Aerodyn.}, Vol. 114, 2013, pp. 12--17.
\newblock \doi{10.1016/j.jweia.2012.12.015}.

\bibitem[{Velasco et~al.(2017)Velasco, {L{\'{o}}pez Mejia}, and
  La{\'{i}}n}]{Velasco2017}
Velasco, D., {L{\'{o}}pez Mejia}, O., and La{\'{i}}n, S., \enquote{{Numerical
  simulations of active flow control with synthetic jets in a Darrieus
  turbine},} \emph{Renew. Energy}, Vol. 113, 2017, pp. 129--140.
\newblock \doi{10.1016/j.renene.2017.05.075}.

\bibitem[{Greenblatt et~al.(2012)Greenblatt, Schulman, and
  Ben-Harav}]{Greenblatt2012}
Greenblatt, D., Schulman, M., and Ben-Harav, A., \enquote{{Vertical axis wind
  turbine performance enhancement using plasma actuators},} \emph{Renew.
  Energy}, Vol.~37, No.~1, 2012, pp. 345--354.
\newblock \doi{10.1016/j.renene.2011.06.040}.

\bibitem[{Morgulis and Seifert(2016)}]{Morgulis2016}
Morgulis, N., and Seifert, A., \enquote{{Fluidic flow control applied for
  improved performance of Darrieus wind turbines},} \emph{Wind Energy},
  Vol.~19, No.~9, 2016, pp. 1585--1602.
\newblock \doi{10.1002/we.1938}.

\bibitem[{Franck and Breuer(2017)}]{Franck2017}
Franck, J.~A., and Breuer, K.~S., \enquote{{Unsteady high-lift mechanisms from
  heaving flat plate simulations},} \emph{Int. J. Heat Fluid Flow}, Vol.~67,
  2017, pp. 230--239.
\newblock \doi{10.1016/j.ijheatfluidflow.2017.08.012}.

\bibitem[{Ribeiro et~al.(2020)Ribeiro, Frank, and Franck}]{Ribeiro2020}
Ribeiro, B. L.~R., Frank, S.~L., and Franck, J.~A., \enquote{{Vortex dynamics
  and Reynolds number effects of an oscillating hydrofoil in energy harvesting
  mode},} \emph{J. Fluids Struct.}, Vol.~94, 2020, p. 102888.
\newblock \doi{10.1016/j.jfluidstructs.2020.102888}.

\bibitem[{Keisar et~al.(2020)Keisar, {De Troyer}, and Greenblatt}]{Keisar2020}
Keisar, D., {De Troyer}, T., and Greenblatt, D., \enquote{{Atypical
  Aerodynamics of Large Chord/Radius Vertical Axis Wind Turbines},} \emph{AIAA
  SciTech Forum}, 2020, pp. 1--9.
\newblock \doi{10.2514/6.2020-1491}.

\bibitem[{Sch{\"{o}}nborn and Chantzidakis(2007)}]{Schonborn2007}
Sch{\"{o}}nborn, A., and Chantzidakis, M., \enquote{{Development of a hydraulic
  control mechanism for cyclic pitch marine current turbines},} \emph{Renew.
  Energy}, Vol.~32, No.~4, 2007, pp. 662--679.
\newblock \doi{10.1016/j.renene.2006.02.004}.

\bibitem[{Paraschivoiu et~al.(2009)Paraschivoiu, Trifu, and
  Saeed}]{Paraschivoiu2009}
Paraschivoiu, I., Trifu, O., and Saeed, F., \enquote{{H-Darrieus Wind Turbine
  with Blade Pitch Control},} \emph{Int. J. Rotating Mach.}, Vol. 2009, 2009,
  pp. 1--7.
\newblock \doi{10.1155/2009/505343}.

\bibitem[{Lazauskas(1992)}]{Lazauskas1992}
Lazauskas, L., \enquote{{Three Pitch Control Systems for Vertical Axis Wind
  Turbines Compared},} \emph{Wind Eng.}, Vol.~16, No.~5, 1992, pp. 269--282.

\bibitem[{Kirke(2011)}]{Kirke2011a}
Kirke, B.~K., \enquote{{Tests on ducted and bare helical and straight blade
  Darrieus hydrokinetic turbines},} \emph{Renew. Energy}, Vol.~36, No.~11,
  2011, pp. 3013--3022.
\newblock \doi{10.1016/j.renene.2011.03.036}.

\bibitem[{Elkhoury et~al.(2015)Elkhoury, Kiwata, and Aoun}]{Elkhoury2015}
Elkhoury, M., Kiwata, T., and Aoun, E., \enquote{{Experimental and numerical
  investigation of a three-dimensional vertical-axis wind turbine with
  variable-pitch},} \emph{J. Wind Eng. Ind. Aerodyn.}, Vol. 139, 2015, pp.
  111--123.
\newblock \doi{10.1016/j.jweia.2015.01.004}.

\bibitem[{{R{\'{e}}mi Gosselin} et~al.(2013){R{\'{e}}mi Gosselin}, {Matthieu
  Boudreau}, and {Guy Dumas}}]{RemiGosselin2013}
{R{\'{e}}mi Gosselin}, {Matthieu Boudreau}, and {Guy Dumas},
  \enquote{{Parametric study of H-Darrieus vertical-axis turbines using uRANS
  simulations},} \emph{21st Annu. Conf. CFD Soc. Canada, Sherbrooke, Qc,
  Canada}, 2013.

\bibitem[{Abdalrahman et~al.(2017)Abdalrahman, Melek, and
  Lien}]{Abdalrahman2017}
Abdalrahman, G., Melek, W., and Lien, F.-S., \enquote{{Pitch angle control for
  a small-scale Darrieus vertical axis wind turbine with straight blades
  (H-Type VAWT)},} \emph{Renew. Energy}, Vol. 114, 2017, pp. 1353--1362.
\newblock \doi{10.1016/j.renene.2017.07.068}.

\bibitem[{Paillard et~al.(2015)Paillard, Astolfi, and Hauville}]{Paillard2015}
Paillard, B., Astolfi, J., and Hauville, F., \enquote{{URANSE simulation of an
  active variable-pitch cross-flow Darrieus tidal turbine: Sinusoidal pitch
  function investigation},} \emph{Int. J. Mar. Energy}, Vol.~11, 2015, pp.
  9--26.
\newblock \doi{10.1016/j.ijome.2015.03.001}.

\bibitem[{Strom et~al.(2017)Strom, Brunton, and Polagye}]{Strom2017}
Strom, B., Brunton, S.~L., and Polagye, B., \enquote{{Intracycle angular
  velocity control of cross-flow turbines},} \emph{Nat. Energy}, Vol.~2,
  No.~8, 2017, pp. 1--9.
\newblock \doi{10.1038/nenergy.2017.103}.

\bibitem[{Issa(1986)}]{issa1986solution}
Issa, R.~I., \enquote{Solution of the implicitly discretised fluid flow
  equations by operator-splitting,} \emph{Journal of Computational Physics},
  Vol.~62, No.~1, 1986, pp. 40--65.

\bibitem[{Weller et~al.(1998)Weller, Tabor, Jasak, and
  Fureby}]{weller1998tensorial}
Weller, H.~G., Tabor, G., Jasak, H., and Fureby, C., \enquote{A tensorial
  approach to computational continuum mechanics using object-oriented
  techniques,} \emph{Computers in Physics}, Vol.~12, No.~6, 1998, pp. 620--631.

\bibitem[{Menter(1994)}]{menter1994two}
Menter, F.~R., \enquote{Two-equation eddy-viscosity turbulence models for
  engineering applications,} \emph{AIAA Journal}, Vol.~32, No.~8, 1994, pp.
  1598--1605.

\bibitem[{Balduzzi et~al.(2016)Balduzzi, Bianchini, Maleci, Ferrara, and
  Ferrari}]{Balduzzi2016}
Balduzzi, F., Bianchini, A., Maleci, R., Ferrara, G., and Ferrari, L.,
  \enquote{{Critical issues in the CFD simulation of Darrieus wind turbines},}
  \emph{Renew. Energy}, Vol.~85, 2016, pp. 419--435.
\newblock \doi{10.1016/j.renene.2015.06.048}.

\bibitem[{Wong et~al.(2018)Wong, Chong, Sukiman, Shiah, Poh, Sopian, and
  Wang}]{Wong2018}
Wong, K.~H., Chong, W.~T., Sukiman, N.~L., Shiah, Y.-C., Poh, S.~C., Sopian,
  K., and Wang, W.-C., \enquote{{Experimental and simulation investigation into
  the effects of a flat plate deflector on vertical axis wind turbine},}
  \emph{Energy Convers. Manag.}, Vol. 160, 2018, pp. 109--125.
\newblock \doi{10.1016/j.enconman.2018.01.029}.

\bibitem[{Rogowski et~al.(2018)Rogowski, Hansen, and
  Maro{\'{n}}ski}]{Rogowski2018}
Rogowski, K., Hansen, M. O.~L., and Maro{\'{n}}ski, R., \enquote{{Steady and
  unsteady analysis of NACA 0018 airfoil in vertical-axis wind turbine},}
  \emph{J. Theor. Appl. Mech.}, Vol.~56, No.~1, 2018, p. 203.
\newblock \doi{10.15632/jtam-pl.56.1.203}.

\bibitem[{Castelli et~al.(2010)Castelli, Ardizzon, Battisti, Benini, and
  Pavesi}]{Castelli2010}
Castelli, M.~R., Ardizzon, G., Battisti, L., Benini, E., and Pavesi, G.,
  \enquote{{Modeling strategy and numerical validation for a Darrieus vertical
  axis micro-wind turbine},} \emph{ASME Int. Mech. Eng. Congr. Expo. Proc.},
  2010, pp. 1--10.
\newblock \doi{10.1115/imece2010-39548}.

\bibitem[{Bardina et~al.(1997)Bardina, Huang, and
  Coakley}]{bardina1997turbulence}
Bardina, J.~E., Huang, P.~G., and Coakley, T.~J., \enquote{Turbulence modeling
  validation, testing, and development,} Tech. rep., NASA Ames Research Center,
  1997.

\bibitem[{Li et~al.(2013)Li, Zhu, Xu, and Xiao}]{Li2013}
Li, C., Zhu, S., Xu, Y.-l., and Xiao, Y., \enquote{{2.5D large eddy simulation
  of vertical axis wind turbine in consideration of high angle of attack
  flow},} \emph{Renew. Energy}, Vol.~51, 2013, pp. 317--330.
\newblock \doi{10.1016/j.renene.2012.09.011}.

\bibitem[{Bianchini et~al.(2017)Bianchini, Balduzzi, Bachant, Ferrara, and
  Ferrari}]{Bianchini2017}
Bianchini, A., Balduzzi, F., Bachant, P., Ferrara, G., and Ferrari, L.,
  \enquote{{Effectiveness of two-dimensional CFD simulations for Darrieus
  VAWTs: a combined numerical and experimental assessment},} \emph{Energy
  Convers. Manag.}, Vol. 136, 2017, pp. 318--328.
\newblock \doi{10.1016/j.enconman.2017.01.026}.

\bibitem[{Snortland et~al.(2019)Snortland, Polagye, and
  Williams}]{Snortland2019}
Snortland, A., Polagye, B., and Williams, O., \enquote{{Influence of Near-blade
  hydrodynamics on Cross-flow Turbine performance},} \emph{Proc. 13th Eur. Wave
  Tidal Energy Conf.}, 2019, pp. 1--9.

\bibitem[{Kinsey and Dumas(2017)}]{Kinsey2017}
Kinsey, T., and Dumas, G., \enquote{{Impact of channel blockage on the
  performance of axial and cross-flow hydrokinetic turbines},} \emph{Renew.
  Energy}, Vol. 103, 2017, pp. 239--254.
\newblock \doi{10.1016/j.renene.2016.11.021}.

\bibitem[{Menter and Esch(2001)}]{Menter2001}
Menter, F., and Esch, T., \enquote{{Elements of Industrial Heat Transfer
  Predictions},} \emph{16th Brazilian Congr. Mech. Eng.}, 2001, pp. 117--127.

\bibitem[{Barnsley and Wellicome(1990)}]{barnsley1990final}
Barnsley, M., and Wellicome, J., \enquote{Final report on the 2nd phase of
  development and testing of a horizontal axis wind turbine test rig for the
  investigation of stall regulation aerodynamics,} Tech. rep., ETSU, 1990.

\bibitem[{Ross and Polagye(2020)}]{Ross2020}
Ross, H., and Polagye, B., \enquote{{An experimental assessment of analytical
  blockage corrections for turbines},} \emph{Renew. Energy}, Vol. 152, 2020,
  pp. 1328--1341.
\newblock \doi{10.1016/j.renene.2020.01.135}.

\bibitem[{Polagye et~al.(2019)Polagye, Strom, Ross, Forbush, and
  Cavagnaro}]{Polagye2019}
Polagye, B., Strom, B., Ross, H., Forbush, D., and Cavagnaro, R.~J.,
  \enquote{{Comparison of cross-flow turbine performance under torque-regulated
  and speed-regulated control},} \emph{J. Renew. Sustain. Energy}, Vol.~11,
  No.~4, 2019, p. 044501.
\newblock \doi{10.1063/1.5087476}.

\bibitem[{Strom et~al.(2018)Strom, Johnson, and Polagye}]{Strom2018}
Strom, B., Johnson, N., and Polagye, B., \enquote{{Impact of blade mounting
  structures on cross-flow turbine performance},} \emph{J. Renew. Sustain.
  Energy}, Vol.~10, No.~3, 2018, p. 034504.
\newblock \doi{10.1063/1.5025322}.

\bibitem[{Zhou et~al.(1999)Zhou, Adrian, Balachandar, and Kendall}]{ZHOU1999}
Zhou, J., Adrian, R.~J., Balachandar, S., and Kendall, T.~M.,
  \enquote{{Mechanisms for generating coherent packets of hairpin vortices in
  channel flow},} \emph{J. Fluid Mech.}, Vol. 387, 1999, pp. 353--396.
\newblock \doi{10.1017/S002211209900467X}.

\bibitem[{Adrian et~al.(2000)Adrian, Christensen, and Liu}]{Adrian2000}
Adrian, R.~J., Christensen, K.~T., and Liu, Z.-C., \enquote{{Analysis and
  interpretation of instantaneous turbulent velocity fields},} \emph{Exp.
  Fluids}, Vol.~29, No.~3, 2000, pp. 275--290.
\newblock \doi{10.1007/s003489900087}.

\end{thebibliography}

\end{document}